\documentclass[fleqn,usenatbib]{mnras}

\usepackage{newtxtext,newtxmath}

\usepackage[T1]{fontenc}
\usepackage{ae,aecompl}

\usepackage{graphicx}	
\usepackage{hyperref}
\usepackage{url}
\usepackage[dvipsnames]{xcolor}
\usepackage{amsmath} 
\usepackage{siunitx}

\newcommand{\gaia}{\textit{Gaia}}
\newcommand{\Teff}{\mbox{$T_{\rm eff}$}}
\newcommand{\Msun}{\mbox{M$_\odot$}}

\newcommand{\logg}{\mbox{$\log(g)$}}

\title[The 40\,pc white dwarf sample]{The 40\,pc sample of white dwarfs from \gaia}

\author[O'Brien et al.]{Mairi W. O'Brien,$^{1}$\thanks{E-mail: mairi.obrien1@gmail.com}
P.-E. Tremblay,$^{1}$ B.~L.~Klein,$^{2}$ D.~Koester,$^{3}$ C.~Melis,$^{4}$
\newauthor A.~B{\'e}dard,$^{1}$ E.~Cukanovaite,$^{1}$ T.~Cunningham,$^{1}$ A.~E.~Doyle,$^{5}$ B.~T.~G\"ansicke,$^{1}$
\newauthor N.~P.~Gentile Fusillo,$^{6}$ M.~A.~Hollands,$^{7}$ J.~McCleery,$^{1}$ I.~Pelisoli,$^{1}$ S.~Toonen,$^{8}$ 
\newauthor A.~J.~Weinberger,$^{9}$ and B.~Zuckerman$^{2}$
  \\
$^{1}$ Department of Physics, University of Warwick, Coventry CV4 7AL, UK \\
$^{2}$ Department of Physics and Astronomy, University of California, Los Angeles, CA 90095-1562, USA \\
$^{3}$ Institut f\"ur Theoretische Physik und Astrophysik, University of Kiel, 24098 Kiel, Germany \\
$^{4}$ Center for Astrophysics and Space Sciences, University of California, San Diego, CA 92093-0424, USA \\
$^{5}$ Department of Earth, Planetary, and Space Sciences, University of California, Los Angeles, CA 90095, USA \\
$^{6}$ Department of Physics and Astronomy, University of Sheffield, Sheffield, S3 7RH, UK \\
$^{7}$ Department of Physics, Universit\`a degli Studi di Trieste, 34127 Trieste, Italy \\
$^{8}$ Anton Pannekoek Institute for Astronomy, University of Amsterdam, 1090 GE Amsterdam, The Netherlands \\
$^{9}$ Earth and Planets Laboratory, Carnegie Institution for Science, 5241 Broad Branch Rd NW, Washington, DC 20015, USA \\
}

\date{Accepted 2023 December 04. Received 2023 November 15; in original form 2023 September 11}

\pubyear{2024}
\begin{document}
\label{firstpage}
\pagerange{\pageref{firstpage}--\pageref{lastpage}}
\maketitle

\begin{abstract}
We present a comprehensive overview of a volume-complete sample of white dwarfs located within 40\,pc of the Sun, a significant proportion of which were detected in \gaia\ Data Release 3 (DR3). Our DR3 sample contains 1076 spectroscopically confirmed white dwarfs, with just five candidates within the volume remaining unconfirmed ($>$\,99\,per\,cent spectroscopic completeness). Additionally, 28 white dwarfs were not in our initial selection from \gaia\ DR3, most of which are in unresolved binaries. We use \gaia\ DR3 photometry and astrometry to determine a uniform set of white dwarf parameters, including mass, effective temperature, and cooling age. We assess the demographics of the 40\,pc sample, specifically magnetic fields, binarity, space density and mass distributions.

\end{abstract}


\section{Introduction}
\label{sec:intro}

Stars with masses below $\approx$ 10\,M$_{\odot}$ will eventually end their lives as white dwarfs. These stellar remnants have no hydrogen or helium left to burn in their cores, so they cool down for the remainder of their lifetimes. Therefore, the temperature of a white dwarf, which is determined using photometry or spectroscopy, is a proxy for the star's cooling age \citep{Hansen1999,Fontaine2001,Althaus2010,Salaris2010,Bedard2020}. Total age and therefore stellar formation history can then be reconstructed from the relation between initial stellar mass and white dwarf final mass \citep{Weidemann1983,Catalan2008,Williams2009,El-Badry2018,Cummings2018}.

Some of the brightest and closest white dwarfs to the Sun were identified and observed spectroscopically as long ago as the 1910s. However, the white dwarf luminosity function peaks at faint magnitudes, and therefore most white dwarfs are cool and faint. The lack of precise parallax measurements for such faint stars meant that identifying nearby white dwarfs has been historically challenging. Within the last two decades, many local white dwarfs have been discovered through spectroscopic and photometric observations \citep[][and others]{Bergeron1997,Bergeron2001,Liebert2005,Kawka2006,Gianninas2011,Sayres2012,Kawka2012}. The Research Consortium on Nearby Stars (RECONS) produced a dedicated series of explicit searches for local white dwarfs based on parallax measurements \citep{Subasavage2007,Subasavage2008,Subasavage2009,Subasavage2017}. Studies of the local white dwarf population within 13\,pc, 20\,pc and 25\,pc volumes were carried out by \citet{Holberg2002,Holberg2008,Holberg2016} and \citet{Giammichele2012}. The first dedicated effort to identify white dwarfs within a 40\,pc volume by \citet{Limoges2015} was limited to the northern hemisphere, and resulted in 281 new discoveries for a total of 492 white dwarfs estimated to be within 40\,pc. 

Following \gaia\ DR2, the first all-sky \gaia\ catalogues of white dwarf candidates with precise parallaxes were released \citep{Jimenez2018,Gentile2019}. \citet{Hollands2018_Gaia} compiled and analysed the volume-complete 20\,pc white dwarf sample from \gaia\ DR2. \gaia\ DR3 has further improved our understanding of the local stellar population within 100\,pc, which is nearly volume-complete \citep{Gaia-nearby-stars,Gentile2021,Jimenez2023}. Meanwhile, almost every white dwarf candidate within the smaller 40\,pc volume has now been confirmed with optical spectroscopy at medium resolution ($R>1000$) and high signal-to-noise (S/N) of $>30$. 

\citet{Tremblay2020} used the \gaia\ DR2 white dwarf catalogue from \citet{Gentile2019} to confirm 179 new white dwarfs within 40\,pc, mostly in the northern hemisphere. \citet{OBrien2023} used the updated \gaia\ DR3 white dwarf catalogue from \citet{Gentile2021} to confirm 203 new white dwarfs within 40\,pc, mostly in the southern hemisphere. With the additional 15 new observations presented in this work, the nature of 1078 \gaia\ white dwarf candidates out of the 1083 from the \citet{Gentile2021} white dwarf catalogue have now been spectroscopically confirmed within 40\,pc. Just two of these are main-sequence contaminants \citep{OBrien2023} and the other 1076 are white dwarfs. Therefore, the \gaia\ DR3 40\,pc white dwarf sample now has 99.3\,per\,cent spectroscopic completeness. In addition, the completeness of the \gaia\ DR3 white dwarf selection from \citet{Gentile2021} at 40\,pc is estimated to be $\approx$ 97\,per\,cent based on pre-\gaia\ surveys and population synthesis \citep[][and this work]{Toonen2017,Hollands2018_Gaia,Mccleery2020}. The 40\,pc white dwarf sample is the largest ever volume-limited sample of white dwarfs with medium-resolution optical spectroscopic follow-up. As noted by \citet{Gentile2021}, reddening effects for white dwarfs closer than 40\,pc are essentially negligible, and therefore no correction is made for reddening in this work.

White dwarf volume samples have been found to have several practical advantages for deriving astrophysical relations, despite suggestions that volume samples reflect a highly sub-optimal selection function \citep{Rix2021}. First and foremost, decades of spectroscopic and spectropolarimetric follow-up work for nearby white dwarfs allows the derivation of stellar parameters that are more precise and accurate than white dwarfs at larger distances \citep{Bergeron2019,Mccleery2020,Bagnulo2022}. Furthermore, white dwarfs with cooling ages $>5$\,Gyr rapidly become fainter than the \gaia\ magnitude limit at distances larger than 40--100\,pc, resulting in increasingly age- and mass-biased samples. Older and heavier white dwarfs that have long cooling ages and short main-sequence lifetimes are intrinsically faint and only seen in the local volume, yet those provide a robust test of old planetary systems \citep{Hollands2018_dz,Kaiser2021,Hollands2021,Blouin2022b,Elms2022} and stellar evolution models, e.g. using wide binaries \citep{El-Badry2021,Qiu2021,Heintz2022,Moss2022}.

The star formation history of the disc of the Milky Way has been determined using white dwarfs \citep[][and references therein]{Fantin2019}. Recently, \citet{Cukanovaite2023} derived the star formation history of the Galactic disc from the 40\,pc white dwarf sample. They find that a uniform stellar formation history with one galactic component provides a good fit to the \gaia\ G-magnitude distribution of the white dwarfs in this local volume. Local white dwarf samples have been used to study the evolution of magnetism in stars and the origin of magnetic white dwarfs \citep{Ferrario2020,Bagnulo2020,Bagnulo2022,Hardy2023}, the initial-final mass relation \citep{El-Badry2018}, core crystallisation \citep{Tremblay2019b,Cheng2019,Mccleery2020,Kilic2020_100pc,Blouin2021}, white dwarf spectral evolution, convective mixing and carbon dredge-up \citep{Blouin2019b,Ourique2020,Cunningham2020,Lopez2022,Blouin2023,Blouin2023b, Camisassa2023}, and binary evolution and gravitational wave background predictions \citep{Toonen2017,Hollands2018_Gaia,Rebassa2021,Torres2022,Korol2022,Kupfer2023}.

With the updated \gaia\ DR3, and the improvement in spectroscopic completeness of the southern 40\,pc white dwarf candidates from \citet{OBrien2023}, we now present a study of the full, unbiased sample of white dwarfs within 40\,pc of the Sun. This follows upon \citet{Mccleery2020} who previously used \gaia\ DR2 to study the 40\,pc northern hemisphere sample. Larger 100\,pc volume samples have also been studied, including the white dwarf sample in the SDSS footprint \citep{Kilic2020_100pc,Caron2023}, and the \gaia\ DR3 sample of low resolution spectra \citep{Jimenez2023,Garcia-Zamora2023,Vincent2023}. However, these samples have a significantly lower volume completeness than the present 40\,pc sample in terms of high S/N and medium resolution spectroscopy. 

Section~\ref{sec:sample} describes the 40\,pc white dwarf sample, considering the \gaia-identified white dwarfs as well as those that are not in the \citet{Gentile2021} \gaia-based catalogue. Section~\ref{sec:disc} discusses aspects of the 40\,pc white dwarf sample, including binarity, magnetism, pollution from planetary debris, and space density. We conclude in Section~\ref{sec:conclusions}.

\section{The 40 pc Sample}
\label{sec:sample}

The 40\,pc white dwarf sample, as discussed in this work, refers to all white dwarf candidates from the \citet{Gentile2021} catalogue, selected from \gaia\ DR3, within 40\,pc of the Sun that have been spectroscopically confirmed. The \gaia\ 40\,pc white dwarf sample, with 1076 members, is listed in the online material (Table~\ref{tab:all_online}). A description of the contents of the online material is in Table~\ref{tab:online_table}. We follow a similar format as the table A1 from \citet{Mccleery2020}. Unless specified, our analysis in this work only considers the white dwarfs in this main \gaia-identified sample \citep{Gentile2021}. We adopt the WD\,Jhhmmss.ss\,$\pm$\,ddmmss.ss naming convention introduced in \citet{Gentile2019}, and we also use the shorthand notation: WD\,Jhhmm\,$\pm$\,ddmm for simplicity. We make a noise cut in \gaia\ DR3, and find $\approx$18\,000 sources within 40\,pc. We select sources with \texttt{parallax\_over\_error} $>$ 1 and outside the Galactic plane \texttt{astrometric\_excess\_noise} $<$ 1.5 but within the Galactic plane \texttt{astrometric\_excess\_noise} $<$ 1. Therefore, we note that about 6\,per\,cent of stars in the local volume are white dwarfs. This result is consistent with the complete RECONS 10\,pc sample, for which 6\,per\,cent (19 out of 316) stars are white dwarfs \citep{Henry2018}.

There are an additional five sources which are white dwarf candidates without spectroscopic follow-up within 40\,pc; these are listed in Table~\ref{tab:WDsnospectra}. There are 15 confirmed and candidate white dwarfs within 1$\sigma_\varpi$ of 40\,pc, listed in Table~\ref{tab:onesigma}. There are 28 known white dwarfs that did not make the cut of \citet{Gentile2021}, mostly due to photometric contamination from nearby bright stars. These are listed in Table~\ref{tab:missingWDs}. In Sections~\ref{sec:binaries} and \ref{sec:space_density}, we specify that we will be including all white dwarfs and candidates from these tables alongside the main 40\,pc white dwarf sample, for completeness in our analysis. 

\begin{figure}
    \centering
	\includegraphics[width=\columnwidth]{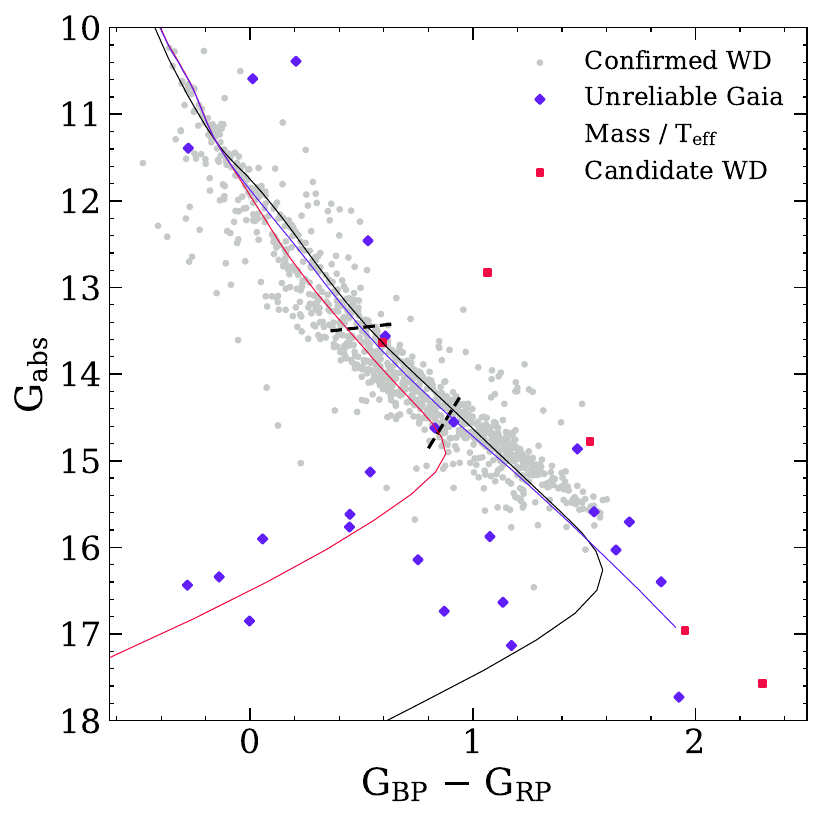}
	\caption{A \gaia\ HR diagram showing all spectroscopically confirmed white dwarfs within 40\,pc that are in the \citet{Gentile2021} catalogue. Confirmed white dwarfs that have unreliable \gaia\ masses determined from \gaia\ photometry and astrometry are shown as purple diamonds. The remaining unobserved candidates from \citet{Gentile2021} are shown as red squares. The purple line indicates pure-He cooling tracks, the black line indicates pure-H cooling tracks and the red line indicates mixed H/He = $10^{-5}$ cooling tracks for a 0.6\,\Msun\ white dwarf. The black dashed lines indicate where 7000\,K (upper line) and 5000\,K (lower line) white dwarfs sit on the cooling tracks.}
    \label{fig:candidatesHR}
\end{figure}

There are also more general issues with the \gaia\ photometric parameters caused by the low-mass issue in models, and which we correct for in our Table~\ref{tab:all_online} and discuss in Section~\ref{sec:corrections}. For our main \gaia-defined 40\,pc sample, there are some white dwarfs which are very cool or have contaminated photometry for which we do not include the effective temperature (\Teff) and surface gravity (\logg) determined from fitting \gaia\ photometry in Table~\ref{tab:all_online}. These white dwarfs are listed in Table~\ref{tab:bad_params}, and we discuss the reasons for these issues in Section~\ref{sec:unreliable_gaia_params}. Fig.~\ref{fig:candidatesHR} shows a \gaia\ Hertzsprung-Russell (HR) diagram of the full 40\,pc sample, including evolution models, where candidate white dwarfs and those with unreliable \gaia\ parameters are highlighted.

The core of a white dwarf is usually comprised of a combination of carbon and oxygen, or oxygen and neon, surrounded by an envelope of helium and hydrogen. It will also have a thin atmosphere dominated by either hydrogen or helium and occasionally contaminated by trace elements. The spectral type of a white dwarf is determined from the strength of atomic lines from optical spectroscopy \citep{Sion1983}. When hydrogen lines are the strongest optical features, it is classified as a DA. Neutral helium gives the classification DB, and a DO implies the detection of ionised helium lines. DQ white dwarfs have detectable carbon, which is typically dredged up from the stellar interior or brought to the surface during a merger (hot DQ), and DZ white dwarfs display metal pollution which is generally attributed to planetary debris. A DC white dwarf has a completely featureless spectrum. Combinations of these letters in an overall spectral type can form a picture of the white dwarf's atmospheric composition, however they may not reflect the relative abundances of each element, and depend on S/N and resolution of the spectra \citep{Doyle2023}. Additionally, below 10\,500\,K helium lines are no longer visible, and below 5000\,K hydrogen lines also disappear in a white dwarf spectrum. Therefore the atmospheric composition for a white dwarf cooler than 5000\,K, where both hydrogen and helium lines are absent, is generally uncertain. 

Every white dwarf in our sample is classified into a spectral type with a published reference (column 21 in Table~\ref{tab:all_online}) which, coupled with a careful inspection of prior spectral modelling \citep[see, e.g.,][]{Limoges2015,Tremblay2020,OBrien2023} and a comparison with \gaia\ photometric parameters \citep{Gentile2021}, informs the atmospheric composition (H- or He-rich; column 12). We use this composition to identify the appropriate set of \gaia\ photometric parameters determined by \citet{Gentile2021}. In Table~\ref{tab:spectraltypes} we show the adopted atmospheric composition for every spectral type in the 40\,pc white dwarf sample. 

We provide a wide binary catalogue in the online material (Table~\ref{tab:widebinaries_online}), for which we show all known 40\,pc wide multiple-star systems where at least one component is a white dwarf. Multiple-star systems are discussed in detail in Section~\ref{sec:binaries}.

We adopt an atmospheric composition of $\log({\rm H/He}) = -5$ (in number of atoms) for \Teff\ $>$ 7000\,K white dwarfs with He-dominated atmospheres, as a pure-He composition does not reproduce the B-branch bifurcation seen in the \gaia\ HR diagram for white dwarfs above this \Teff\ \citep{Bergeron2019}. The H in this composition is typically below the optical spectroscopic detection limit. \citet{Blouin2023} and \citet{Camisassa2023} have demonstrated that an atmospheric composition of He with trace C below the optical detection limit better reproduces the bifurcation. Both trace C and H contribute to increase the number of free electrons, and hence the strength of the He$^{-}$ free-free opacity at optical wavelengths (e.g. \citealt{Provencal2002}). Therefore, the effects of C and H are fully degenerate for individual white dwarfs, unless detailed abundances are available, e.g. C/He for DQ stars, or H/He for a handful of He-rich DA or DZA. As a consequence, we continue to adopt the H/He mixed model atmospheres \citep{Mccleery2020} to account for the effect of both H and C in He-rich atmospheres, where $\log({\rm H/He}) = -5$ reproduces the median position of the B-branch. We opt to not use models with tailored atmospheric compositions for DZ and DQ white dwarfs for the reasons explained above and homogeneity, as these models change the \Teff\ and mass by less than 3\% and 0.05 \Msun, respectively, in the majority of cases. See \citet{Blouin2019c}, \citet{Coutu2019} and \citet{Caron2023} for abundances and parameters of DQ and DZ white dwarfs calculated using tailored models, some of which overlap with the 40 pc sample presented in this work. The presence of large abundances of metals in cool He-dominated atmospheres can have a significant effect on the derived \Teff\ and \logg\ (e.g. \citealt{Dufour2010,Izquierdo2023}).

Below $\approx$ 7000\,K, models with $\log({\rm H/He}) = -5$ composition start to develop bluer colours \citep{Gentile2020,Bergeron2022} due to collision-induced absorption (CIA), which is not observed in the large majority of white dwarfs in the local sample. This could be a consequence of using H as a proxy for C in the model atmospheres, as carbon is not predicted to contribute to any IR opacity. The spectral evolution of trace H and C abundances in He-rich atmospheres at cool temperatures remains only partially understood \citep{Blouin2019c,Bergeron2022}. Hydrogen would need to be almost fully removed from He-rich models to fit most \gaia\ observations, which is at odds with our current understanding of spectral evolution \citep{Blouin2023}, or H$_2$-He CIA opacity could be modified to match the observations \citep{Bergeron2022}. Instead of using ad-hoc corrections, we continue to employ pure-helium models for He-rich atmospheres below 7000\,K, which fit reasonably well to the \gaia\ white dwarf cooling track (Fig.~\ref{fig:candidatesHR}).

Below 5000\,K, the large majority of white dwarfs are of the DC spectral type, and the atmospheric composition is unconstrained from spectroscopy alone. Optical and IR photometry shows a single, homogeneous cool white dwarf population \citep{Gentile2020}, not enabling the separation between H- and He-rich atmospheres from photometry alone. This might be evidence for spectral evolution to H-rich composition for the vast majority of cool white dwarfs \citep{Caron2023}, although there is no direct evidence nor models that predict that spectral evolution takes place in this temperature range. Rather, evidence from the cool DZ population, where the broadening of metal lines depends on atmospheric composition, 
suggests that both H- and He-rich atmospheres are frequent below 5000\,K \citep{Dufour2007,Blouin2018,Hollands2021,Kaiser2021,Elms2022}, plausibly at the He-rich/H-rich frequency of $\approx$ 30\,per\,cent seen in the warmer range 7000\,$\gtrsim$\,\Teff\,$\gtrsim$\,5000\,K \citep{Blouin2019b,Mccleery2020,Lopez2022}. For simplicity, we use pure-H models for all DC white dwarfs with \Teff\,$<$\,5000\,K as both pure-H and pure-He models predict similar \gaia\ fluxes in this range (Fig.~\ref{fig:candidatesHR}).

\begin{table*}
	\centering
        \caption{Format of the online 40\,pc catalogue which is accessible at this \href{https://cygnus.astro.warwick.ac.uk/phrtxn/}{link}.}
        \label{tab:online_table}
        \begin{tabular}{llll}
                \hline
                Index & Column Name & Units & Description \\
                \hline
                1 & WDJ\_name & $-$ & WD\,J (RA) hhmmss.ss\,$\pm$\, (Dec) ddmmss.ss, equinox and epoch 2000 \\
                2 & DR3\_source\_id & $-$ & \gaia\ DR3 source identifier \\
                3 & parallax & mas & \gaia\ DR3 parallax \\
                4 & parallax\_error & mas & \gaia\ DR3 parallax standard error \\
                5 & ra & deg & Right ascension (J2015.5) \\
                6 & ra\_error & deg & Standard error of right ascension \\
                7 & dec & deg & Declination (J2015.5) \\
                8 & dec\_error & deg & Standard error of declination \\
                9 & absG & magnitude & Absolute \gaia\ DR3 G magnitude \\
                10 & bp\_rp & magnitude & \gaia\ DR3 BP minus \gaia\ DR3 RP colour index \\
                11 & SpT & $-$ & Spectral type \\
                12 & comp & $-$ & Atmospheric composition (H for hydrogen-dominated or He for helium-dominated) \\
                13 & gaia\_teff & K & Adopted \gaia\ DR3 effective temperature \\
                14 & gaia\_teff\_err & K & Standard error on adopted \gaia\ DR3 effective temperature \\
                15 & gaia\_logg & cm s$^{-2}$ & Adopted \gaia\ DR3 surface gravity \\
                16 & gaia\_logg\_err & cm s$^{-2}$ & Standard error on adopted \gaia\ DR3 surface gravity \\
                17 & gaia\_mass & \Msun & Adopted \gaia\ DR3 mass \\
                18 & gaia\_mass\_err & \Msun & Standard error on adopted \gaia\ DR3 mass \\
                19 & corrected\_teff & K & Effective temperature after low-mass correction (see Section~\ref{sec:corrections}) \\
                20 & corrected\_mass & \Msun & Mass after low-mass correction (see Section~\ref{sec:corrections}) \\
                21 & corrected\_age & Gyr & Cooling age after low-mass correction (see Section~\ref{sec:corrections}) \\
                22 & bibcode & $-$ & Reference paper for spectral type \\
                23 & comment & $-$ & Additional comment \\
                \hline
        \end{tabular}\\
\end{table*} 

\begin{table*}
	\centering
        \caption{All white dwarf spectral types listed in the 40\,pc sample, where photometric model composition refers to composition-selected \citet{Gentile2021} parameters. Spectral types ending with H (Zeeman splitting) or P (polarised) imply magnetism, which does not impact atmospheric composition. In all cases other than for DA and DAH/P, we add \textit{(H/P)} to spectral types to note that some white dwarfs in that group are magnetic. The lowercase `e' indicates the presence of emission features.}
        \label{tab:spectraltypes}
        \begin{tabular}{lll}
                \hline
                Spectral type & Number in & Photometric model \\
                (SpT) & 40\,pc & composition \\
                \hline
                DA & 538 & pure-H (except for 2 He-rich DA)\\ 
                DAH or DAP & 64 & pure-H \\ 
                DA(H)e & 4 & pure-H \\ 
                DAZ\textit{(H/P)} & 53 & pure-H \\ 
                DB\textit{(H/P)} & 9 & $\log({\rm H/He})$ $= -$5 \\
                DBA\textit{(H/P)} & 6 & $\log({\rm H/He})$ $= -$5 \\
                DBQA & 1 & $\log({\rm H/He})$ $= -$5 \\
                DBZA & 1 & $\log({\rm H/He})$ $= -$5 \\
                DC\textit{(H/P)} & 287 & $\log({\rm H/He})$ $= -$5, pure-He below 7000\,K, assumed pure-H below 5000\,K \\
                DQ\textit{(H/P)} & 34 & $\log({\rm H/He})$ $= -$5, pure-He below 7000\,K \\
                warm DQ & 2 & pure-He \\
                DQpec\textit{(H/P)} & 8 & $\log({\rm H/He})$ $= -$5, pure-He below 7000\,K \\
                DQZ & 3 & $\log({\rm H/He})$ $= -$5, pure-He below 7000\,K \\
                DX\textit{(H/P)} & 4 & dependent on individual atmospheric analysis \\
                DZ\textit{(H/P)} & 45 & $\log({\rm H/He})$ $= -$5, pure-He below 7000\,K, except pure-H for H-rich DZ \\
                DZA\textit{(H/P)} & 14 & $\log({\rm H/He})$ $= -$5, pure-He below 7000\,K, except pure-H for H-rich DZA \\
                DZAB & 1 & $\log({\rm H/He})$ $= -$5, pure-He below 7000\,K \\
                DZQ\textit{(H/P)} & 2 & $\log({\rm H/He})$ $= -$5, pure-He below 7000\,K  \\
                \hline
        \end{tabular}\\
\end{table*}

\subsection{Correction to Mass and Effective Temperature}
\label{sec:corrections}

A population of white dwarfs evolved from single star evolution is expected to have an essentially constant median mass, independent of temperature \citep{Tremblay2016}. This is in contrast with atmosphere-modelled observations, where there is a low-mass issue found when fitting white dwarf optical photometry.

White dwarfs with H-rich and with He-rich atmospheres that are cooler than $\approx$ 6000\,K are found to have significantly lower masses than the canonical $\approx$ 0.6\,\Msun\ value \citep{Hollands2018_Gaia,Bergeron2019,Blouin2019b,Mccleery2020,Hollands2021}, most likely due to inaccuracies of opacities in the atmospheric models \citep{Caron2023}, for example a problem with the red wing of Ly\,$\alpha$. If we assume instead that these cool low-mass white dwarfs are unresolved double white dwarf binaries, the implication of a trend to lower masses as a function of age is that more binaries formed at a certain time in the past and dominate only at a specific temperature or age. This is not consistent with binary evolution theory, and therefore it is more likely that the low-mass trend is caused by incorrect opacities.

\begin{figure*}
    \centering
	\includegraphics[width=\textwidth]{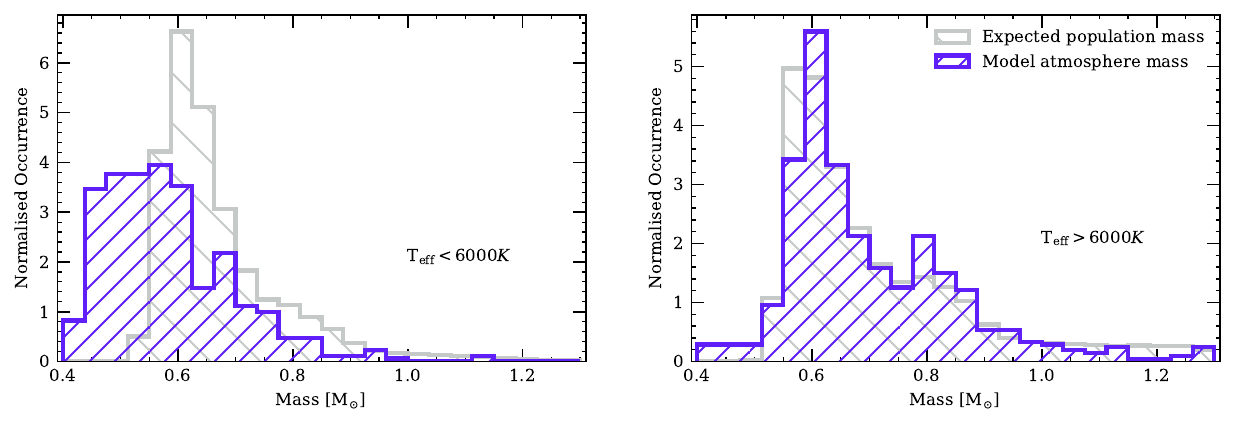}
	\caption{Mass distribution of white dwarfs in the 40\,pc sample determined from \gaia\ photometry. Also shown is the mass distribution of simulated white dwarf masses, with estimated observational errors drawn from a normal distribution with $\mu =$ 0 and $\sigma =$ 0.02\,\Msun, taken from \citet{Cunningham2023}. Double-degenerate candidates have been removed from the observational distribution. Left panel: white dwarfs with \Teff\ below 6000\,K; right panel: white dwarfs with \Teff\ above 6000\,K.}
    \label{fig:mass_distribution_simulation}
\end{figure*}

The low-mass issue is demonstrated in Fig.~\ref{fig:mass_distribution_simulation}, where the 40\,pc mass distribution determined from \gaia\ photometry is compared with synthetic white dwarf masses from \citet{Cunningham2023}. The synthetic mass distribution is formed from a Monte Carlo simulation of a Galactic disc population of main-sequence stars put through an initial-final mass relation based on the 40\,pc sample. There is only a small difference between the synthetic masses for cool and warm white dwarfs, however, in the atmosphere-modelled observations we see a strong excess of lower-mass white dwarfs below 6000\,K.

\subsubsection{Effect of Changing the Ly $\alpha$ Opacity}
\label{sec:lymanalpha}

We test the effect of correcting opacities in the atmosphere models on the masses determined from \gaia\ photometry. The dominant opacities at optical wavelengths (0.3 $-$ 1.0 $\mu$m) for a pure-H atmosphere at $\approx$ 4000\,K \citep[see figure 17 of][]{Saumon2022} are the red wing of Ly\,$\alpha$ \citep{Kowalski2006,Rohrmann2011} and H$^{-}$ bound-free. CIA opacity is dominant in the IR, and hence can also indirectly influence the overall optical flux by energy redistribution. 

\begin{figure}
    \centering
	\includegraphics[width=\columnwidth]{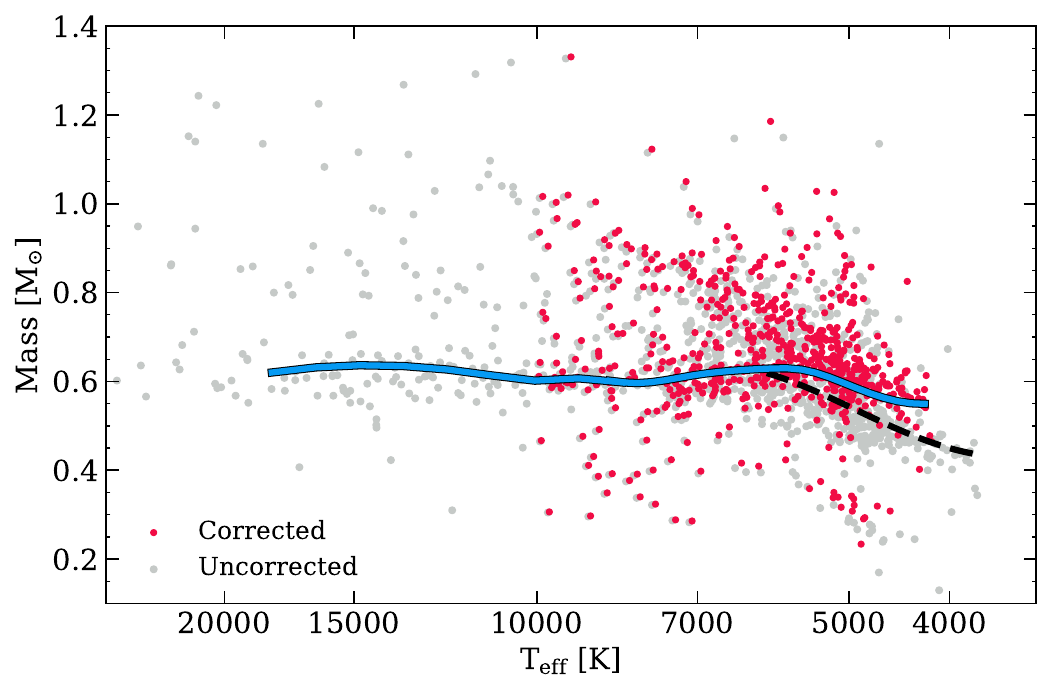}
	\caption{The effect of using pure-H atmosphere models with the Ly$\alpha$ red wing opacity of \citet{Kowalski2006} multiplied by a factor of five on the resulting white dwarf mass and \Teff, for objects with \Teff\ $<$ 10\,000\,K. The dashed black line shows the median mass before the model corrections, and the solid blue line shows the median mass after these corrections. The corrections are only applied to white dwarfs with H-rich or assumed H-rich compositions.}
    \label{fig:massteffcorr_lya}
\end{figure}

We have recomputed our grid of pure-H model atmospheres and spectra by multiplying the overall Ly\,$\alpha$ H-H$_2$ opacity of \citet{Kowalski2006} by an illustrative factor of five. The resulting \gaia\ parameters for white dwarfs with \Teff\ $<$ 10\,000\,K are shown in Fig.\,\ref{fig:massteffcorr_lya}. The results of this change in opacity demonstrate that a possible uncertainty in the strength of this opacity at visible wavelengths is a plausible explanation for the low-mass issue observed in cool white dwarfs in the optical. However, despite this, the median mass following this correction, as shown in Fig.~\ref{fig:massteffcorr_lya} is not constant for cooling white dwarfs and therefore would need further tweaking.

\subsubsection{Ad-hoc Correction}

\begin{figure}
    \centering
	\includegraphics[width=\columnwidth]{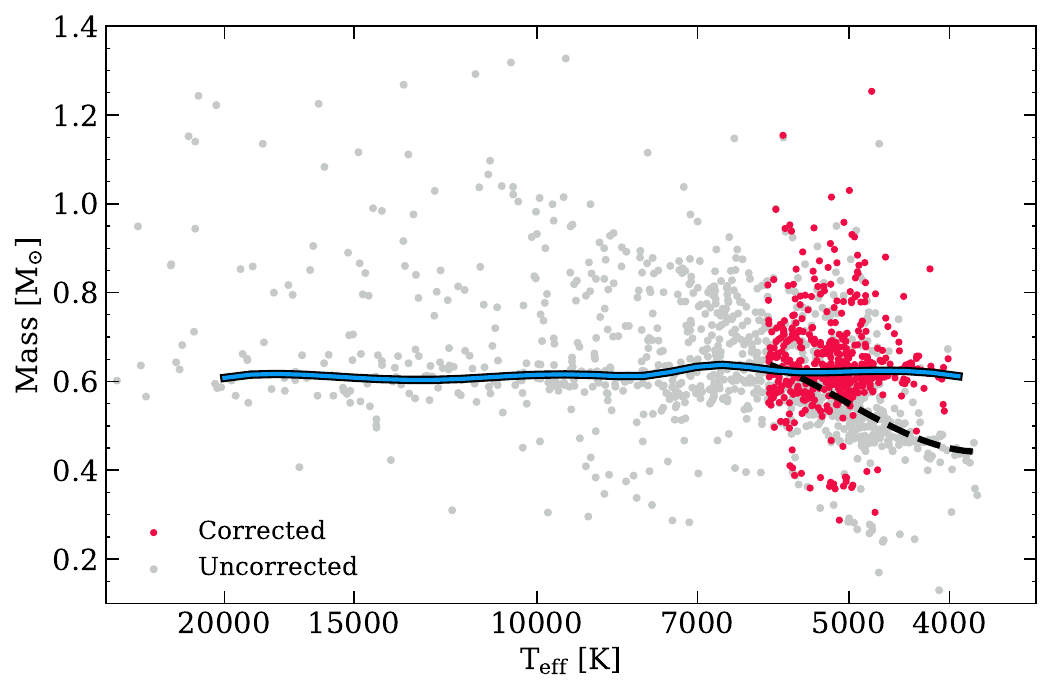}
	\caption{Corrected \gaia\ mass and \Teff\ for all 40\,pc white dwarfs below 6000\,K (red points) with our ad-hoc mass correction, compared to the original mass distribution (gray points). The dashed black line shows the median mass before the mass and \Teff\ corrections, and the solid blue line shows the median mass after these corrections. The median mass in the stable range of 13\,000 $>$ \Teff\ $>$ 8000\,K is used as a reference for the correction.}
    \label{fig:massteffcorr}
\end{figure}

Missing and incorrect opacities in low-\Teff\ and high-pressure white dwarf atmospheres is a major challenge for the research area \citep{Saumon2022}, and it is out of the scope of this work to attempt to solve these issues. Therefore, in order to obtain the expected constant median mass for cooling white dwarfs, we choose to apply an ad-hoc \gaia\ mass and \Teff\ correction on all the white dwarfs in our sample that have an initial \gaia\ \Teff\ less than 6000\,K, the effect of which is shown in Fig.~\ref{fig:massteffcorr}. Correcting the mass and \Teff\ in this way enables us to proceed with meaningful analysis of the volume-complete sample, for which 45\,per\,cent of white dwarfs have \Teff\ $<$ 6000\,K and would otherwise have unreliable masses from photometry.

For our correction, we first fit a function to the median mass distribution of white dwarfs with \Teff\ values below 6000\,K (shown by the black dashed line in Fig.~\ref{fig:massteffcorr}). When calculating median masses as discussed here, we remove double degenerate candidates and those white dwarfs on the crystallisation sequence, to ensure we are correcting to the canonical $\approx$ 0.6\,\Msun\ value. We then fit a correction function so that the median mass as a function of \Teff\ tends towards the canonical mass, which is the median mass in the stable range of 8000 $<$ \Teff\ $<$ 13\,000\,K (shown by the solid blue line in Fig.~\ref{fig:massteffcorr}). We then apply this mass correction to all white dwarfs with \Teff\ $<$ 6000\,K. 

Once the mass has been corrected, the corresponding \Teff\ must be corrected according to the white dwarf mass-radius relation \citep{Bedard2020} and the Stefan-Boltzmann law to reproduce its known luminosity. For all analysis in this paper, \gaia\ mass and \Teff\ have been corrected using this method. Given a mass correction, the radius correction will be roughly the same at all masses up to 1.1\,\Msun, which includes every white dwarf with \Teff\ $<$ 6000\,K in our sample, hence we do not include a mass dependence in the correction.

The function we use to correct the \gaia\ masses for \Teff\ $<$ 6000\,K is a fifth-order polynomial:

\begin{multline}
    \Delta M(T_{\rm eff, i}) = \num{-4.613e-19} T_{\rm eff, i}^5 + \num{1.726e-14} T_{\rm eff, i}^4 \\ - \num{2.486e-10} T_{\rm eff, i}^3 + \num{1.706e-06} T_{\rm eff, i}^2 - 0.005487 T_{\rm eff, i} + 7.068,
    \label{eq:polynomial}
\end{multline}

where $T_{\rm eff, i}$ is the initial uncorrected \Teff. The corrected mass will therefore be:
\begin{equation}
    M_{c} = M_{i} + M_{\rm med} - \Delta M(T_{\rm eff, i}),
    \label{eq:mass_corr}
\end{equation}
{\noindent}where $M_{i}$ is the initial uncorrected mass, $M_{\rm med}$ is the median of the canonical mass in the stable mass range (0.63\,\Msun\ for this sample), and $M_{c}$ is the final corrected mass.

After applying Equations~\ref{eq:polynomial} and \ref{eq:mass_corr}, we then correct the \Teff\ by combining the white dwarf mass-radius relation and the Stefan-Boltzmann law:
\begin{equation}
    \frac{T_{\rm eff, c}}{T_{\rm eff, i}} = \left(\frac{M_{c}}{M_i}\right) ^{1/6}, 
\end{equation}
{\noindent}where $T_{\rm eff, c}$ is the corrected \Teff.

Following our correction, the median mass for standard single white dwarfs (solid blue line in Fig.~\ref{fig:massteffcorr}) is relatively constant as \Teff\ decreases, which is as expected. There is a small increase around 7000\,K due to the overlap of the crystallisation sequence with the canonical-mass white dwarfs. 

Due to their the ad-hoc nature, these corrections are only applicable to masses determined from \gaia\ photometry. The analysis presented in this work relies on \gaia\ mass and \Teff\ values that have been corrected in this way. Columns 19 and 20 of Table~\ref{tab:online_table} correspond to these corrected mass and \Teff\ values respectively, where the statistical uncertainties should be taken to be the same as those without corrections. We also provide mass and \Teff\ values that have not been corrected in Table~\ref{tab:online_table}. Ad-hoc mass corrections have been used in past white dwarf studies, e.g. \citet{Bergeron1994, Giammichele2012}, for other issues that have now been largely resolved with better models \citep{Tremblay2009,Tremblay2013}.

\subsection{White Dwarfs with Unreliable Gaia Masses and Temperatures}
\label{sec:unreliable_gaia_params}

We do not provide \gaia\ \Teff, \logg\ and mass in Table~\ref{tab:all_online} for 25 white dwarfs in the 40\,pc sample. These white dwarfs are listed in Table~\ref{tab:bad_params} and correspond to the purple diamond points in Fig.~\ref{fig:candidatesHR}. For some, their \gaia\ photometry has been contaminated by a bright main-sequence star which is either a companion or a background object. 

Many of the white dwarfs in Table~\ref{tab:bad_params} display signs of collision-induced absorption (CIA) which greatly affects the shape of their spectral energy distributions. White dwarfs displaying strong CIA opacity are classified as IR-faint \citep{Kilic2020_100pc,Bergeron2022,Elms2022}. The parameters listed in Table~\ref{tab:bad_params} for these IR-faint white dwarfs are heavily dependent on the atmosphere models used to fit them, since different codes use different microphysics to account for the extremely dense atmospheres of these stars. Due to their positions on the \gaia\ HR diagram, these white dwarfs are potentially ultra-cool (\Teff\ $<$ 4000\,K). However, \citet{Bergeron2022} suggest that many IR-faint white dwarfs are warmer than 4000\,K (see parameters in Table~\ref{tab:bad_params}). Spectral features caused by metal pollution enable a more accurate determination of parameters in these white dwarfs displaying CIA \citep{Elms2022}, but in most cases their spectra are featureless. 

There are also two white dwarfs in Table~\ref{tab:bad_params} for which strong molecular carbon absorption bands dramatically affect the \gaia\ colours of the white dwarf. Therefore, we do not consider the parameters from \gaia\ photometric fitting of these white dwarfs to be reliable.

\subsection{New Spectroscopic Observations}

We present 16 new white dwarfs within 40\,pc for which there are no previous spectroscopic observations at medium resolution or higher. Their parameters are presented in Table~\ref{tab:WDnew}. We used the following spectroscopic instruments to confirm these new candidates: the High Resolution Echelle Spectrometer (HIRES) on the Keck 10-m telescope \citep{HIRES1994}, the Kast Double Spectrograph on the 3-m Shane telescope, the Magellan Echellete (MagE) and Magellan Inamori Kyocera Echelle (MIKE) instruments on the 6.5-m Magellan telescopes \citep{Magellan2008}, and X-Shooter spectrograph on the VLT \citep{Vernet2011}. We also present new spectra of stars previously confirmed as white dwarfs, where spectral types have been updated following higher-resolution or wider wavelength coverage of the new spectra. 

A cool DQpec and a cool DC observed with X-Shooter are shown in Fig.~\ref{fig:dq_XShooter}. Two DZ(A)(H) white dwarfs observed with HIRES are shown in Fig.~\ref{fig:dz_HIRES_blue}. Hydrogen Balmer lines from DA and DAZ white dwarfs are shown in Figs.~\ref{fig:da_MagE_MIKE_HIRES}$-$\ref{fig:da_HIRES}. Ca\,\textsc{ii} H+K lines in the DAZ white dwarfs are shown in Fig.~\ref{fig:daz_calcium}. All new Kast spectra are shown in Figs.~\ref{fig:kast}$-$\ref{fig:kast2}. Following the addition of these 16 new white dwarfs to our sample, the \gaia\ 40\,pc sample has 1076 confirmed white dwarfs out of 1083 candidates from \citet{Gentile2021}. There are two confirmed main-sequence star contaminants in the sample \citep{OBrien2023}, designated as \textbf{WD\,J092424.45$-$181859.87} and \textbf{WD\,J173230.79$-$171033.14} in \citet{Gentile2021}, which leaves five candidates without spectroscopic follow-up (see Section~\ref{sec:missing} for details). 

\begin{table*}
	\centering
        \caption{White dwarfs within 40\,pc confirmed in this work with spectroscopic follow-up.}
        \label{tab:WDnew}
        \begin{tabular}{llllllll}
                \hline
                WD\,J Name & Parallax & SpT & \gaia\ \Teff & \gaia\ \logg & Instrument & Date of \\
                 & [mas] &  & [K] &  &  & Observation\\
                \hline
                011103.67$-$722741.26 & 34.78 (0.07) & DC & 4160 (130) & 7.72 (0.09) & Magellan/MIKE & 2023/09/21 \\
                $*$021348.83$-$334530.03 & 53.33 (0.06) & DAZ & -- & -- & Magellan/MIKE & 2021/12/19 \\
                023538.55$-$303225.52 & 30.6 (0.2) & DC & -- & -- & VLT/X-Shooter & 2023/08/09 \\
                031330.78$-$424243.22 & 25.73 (0.09) & DC & 4990 (60) & 7.96 (0.05) & Magellan/MIKE & 2023/09/21 \\
                050600.41+590326.89 & 27.7 (0.3) & DC & -- & -- & Shane/Kast & 2023/10/20 \\
                055602.01+135446.71 & 36.53 (0.08) & DA & 5020 (70) & 7.92 (0.05) & Shane/Kast & 2021/09/27 \\
                090834.39+172148.53 & 30.68 (0.06) & DC & 4950 (60) & 7.32 (0.05) & Shane/Kast & 2021/11/13 \\
                102926.67+125733.40 & 27.8 (0.2) & DZH & 5496 (100) & 8.18 (0.07) & Shane/Kast & 2023/05/15 \\
                110143.04+172139.39 & 34.67 (0.05) & DA & 7710 (210) & 8.39 (0.06) & Shane/Kast & 2023/05/15 \\
                $*$115454.07$-$623919.42 & 44.54 (0.05) & DAZ & 4950 (160) & 7.8 (0.1) & Magellan/MagE & 2022/03/23 \\
                115954.88$-$601625.45 & 38.50 (0.06) & DA & 4780 (50) & 7.79 (0.03) & Magellan/MagE & 2022/03/23 \\
                151358.72$-$201445.94 & 36.26 (0.02) & DA & 10900 (110) & 7.98 (0.02) & Keck/HIRES & 2018/05/18 \\
                171409.55$-$053419.96 & 38.20 (0.03) & DA & 9630 (80) & 8.16 (0.02) & Keck/HIRES & 2018/05/18 \\
                171955.76+363936.32 & 28.54 (0.05) & DQpec & 6730 (390)	& 8.4 (0.2) & Shane/Kast & 2023/06/25 \\
                174512.54$-$215309.25 & 25.5 (0.3) & DC & 3980 (240) & 7.6 (0.2) & Shane/Kast & 2023/06/25 \\
                184700.42+181107.49 & 34.49	(0.04) & DA & 8540 (230) & 8.19 (0.06) & Shane/Kast & 2023/05/15 \\
                $*$192743.10$-$035555.23 & 41.93 (0.04) & DZA & 6850 (50) & 8.07 (0.02) & Keck/HIRES & 2019/07/07; 2019/09/07 \\
                 & & & & & Shane/Kast & 2018/08/02; 2019/07/26 \\
                193501.33$-$072527.42 & 24.9 (0.2) [1$\sigma_\varpi$] & DC & 4150 (170) & 7.5 (0.1) & Magellan/MIKE & 2023/09/21 \\
                $*$231732.63$-$460816.77 & 26.0 (0.3) & DQpec & -- & -- & VLT/X-Shooter & 2023/06/20 \\
                \hline
        \end{tabular}\\
        $*$: Spectroscopic \Teff\ and \logg\ are presented in Table~\ref{tab:metal_abundances}.\\
\end{table*} 

\begin{table*}
	\centering
        \caption{Known white dwarfs within 40\,pc with new spectroscopic follow-up and updated spectral types.}
        \label{tab:WDupdated}
        \begin{tabular}{llllllll}
                \hline
                WD\,J Name & Parallax & Old SpT & Updated SpT & \gaia\ \Teff & \gaia\ \logg & Instrument & Date of \\
                 & [mas] & (Reference) & & [K] & & & Observation\\
                \hline
                031907.61+423045.45 & 32.71	(0.03) & DC (1) & DBA & 10\,970 (130) & 8.22 (0.02) & Shane/Kast & 2016/09/23 \\
                131830.01+735318.25 & 27.4 (0.1) & DC: (2) & DA & 5000 (40) & 7.35 (0.04) & Shane/Kast & 2022/04/09 \\
                191936.23+452743.55 & 35.70	(0.04) & DC: (2) & DA & 4780 (20) & 7.31 (0.02) & Shane/Kast & 2021/11/14  \\
                $*$214157.57$-$330029.80 & 62.07 (0.02) & DZH (3) & DZAH & 7110 (50) & 8.00 (0.02) & Keck/HIRES & 2008/08/06; 2008/08/07; \\
                 & & & & & & & 2008/11/14 \\
                223607.66$-$014059.65 & 25.63 (0.04) & DAH: (2) & DAH & 10\,020 (160) & 8.37 (0.03) & Shane/Kast & 2018/07/16; 2016/09/22 \\
                \hline
        \end{tabular}\\
        Notes: (1) \citet{Tremblay2020}, (2) \citet{OBrien2023}, (3) \citet{Bagnulo2019}. $*$: Spectroscopic \Teff\ and \logg\ are presented in Table~\ref{tab:metal_abundances}.\\
\end{table*} 

We determine equivalent widths of the Ca\,\textsc{ii} K lines in \textbf{WD\,J0213$-$3345} and \textbf{WD\,J1154$-$6239}, as 450\,m\AA\ and 220\,m\AA\ respectively. We present best-fit results from the combined spectra and available photometry alongside metal abundances for the DAZ white dwarfs in Table~\ref{tab:metal_abundances}. These Ca\,\textsc{ii} lines are not likely to be interstellar in origin, given that in both cases the radial velocities of the lines are in agreement with the photospheric velocity as best as it can be determined from H$\alpha$. 

\textbf{WD\,J0213$-$3345} had moved to within 1.1\,arcsec of an equally bright star during the epoch of the \gaia\ observations \citep{Hollands2018_Gaia}. In \gaia\ DR3, this star has a renormalised unit weight error (RUWE) of 1.8. Therefore, the \gaia\ photometry is likely contaminated. In our combined fit, we incorporate APASS photometry from 2012 and 2MASS photometry from 2003, which are likely to be less contaminated given the proper motion of 0.4\,arcsec/year. 

\textbf{WD\,J2236$-$0140} was theorised to be a highly magnetic DAH white dwarf in \citet{OBrien2023}, but due to the limited resolution and coverage of the available spectrum, its field strength could not be constrained. Using the Kast spectrum presented in Fig.~\ref{fig:kast}, we confirm that this white dwarf is indeed a high-field DAH.

\textbf{WD\,J2317$-$4608} is a DQpec white dwarf with strong carbon features (see Fig.~\ref{fig:dq_XShooter}). It has a wide main-sequence companion separated by 330\,au (6\,arcsec on-sky separation), which has contaminated the infrared photometry of the white dwarf, and the \gaia\ RP colour is also potentially affected by the companion. For this reason, we fix the \logg\ and determine the \Teff\, (4075\,K) and carbon abundance (see Table~\ref{tab:metal_abundances}). Our models do not account for the distortions of the carbon Swan bands, which is associated with the DQpec class. Therefore the model plotted in Fig.~\ref{fig:dq_XShooter} does not accurately trace the carbon features. With an absolute \gaia\ G value of this white dwarf of 16.40, this star is significantly fainter than any DQ in the Montreal White Dwarf Database \citep{Dufour2017}, and therefore is potentially the coolest confirmed DQ white dwarf with a calculated carbon abundance. 

\textbf{WD\,J0235$-$3032} is an IR-faint DC white dwarf that displays strong signs of CIA (see Fig.~\ref{fig:dq_XShooter}). Only \gaia\ and Pan-STARRS photometry are available for this white dwarf, so we cannot accurately constrain its \Teff, mass and atmospheric composition without near-IR photometry. Similarly, \textbf{WD\,J0506+5903} is also a very blue, IR-faint white dwarf (see Fig.~\ref{fig:kast2}).

\textbf{WD\,J2141$-$3300} and \textbf{WD\,J1927$-$0355} are highly metal-polluted white dwarfs with He-dominated atmospheres. The \gaia\ photometry of these stars indicates \Teff\ $\approx$ 7000\,K and \logg\ $=$ 8 for both objects, assuming no metals. However, the spectra show very strong features that influence the photometry. \textbf{WD\,J2141$-$3300} is commonly known as \textbf{WD\,2138$-$332} and was discovered as a polluted white dwarf by \citet{Subasavage2007}. \textbf{WD\,J1927$-$0355} was first identified as polluted by the Kast spectrograph, and was followed up with the HIRES instrument. 

The HIRES data of \textbf{WD\,J2141$-$3300} were reduced using \texttt{PyRAF} following \citet{Klein2010}, and those for \textbf{WD\,J1927$-$0355} were reduced with the \href{https://sites.astro.caltech.edu/~tb/makee/}{MAKEE} pipeline. Both stars are cool DZs with very strong and broad features, especially blueward of 4000\,\AA, where there is nowhere any continuum. Thus for the echelle spectra we followed \citet{Klein2011}, as well as visually comparing with lower resolution spectra to align, trim, and merge the echelle orders to obtain a continuous output spectrum. In particular, continuum fits of the calibration stars BD+28\,4211, G191$-$B2B, and EGGR\,131, were used to model and divide out the instrumental response function. The overall continuum levels and slopes resulting from these procedures are not absolutely calibrated, so we applied large-scale adjustments to absolute flux levels and slopes to improve display with the models. In our best-fit models, we have set the hydrogen abundance $\log({\rm H/He})$ $= -$3.5. The parameters and metal abundances for these white dwarfs are provided in Table~\ref{tab:metal_abundances}. Both the spectra and best-fit models are shown in Fig.~\ref{fig:dz_HIRES_blue}. Further analysis of the accreted material observed in these two stars will be presented in a future paper.

\textbf{WD\,J1719+3639} was observed with the Kast spectrograph (see Fig.~\ref{fig:kast2}). It appears to show features resembling carbon Swan bands. The spectrum of \textbf{WD\,J1719+3639} is similar to that of \textbf{SDSS\,J161847.38+061155.2}, which was designated as a problematic DQpec object by \citet{Blouin2019c}. We also tentatively classify this white dwarf as a DQpec. \textbf{WD\,J1029+1257} was also observed with the Kast spectrograph and is shown in Fig.~\ref{fig:kast2}. This star has distorted Ca\,\textsc{ii} H+K-lines that are indicative of a magnetic field, and therefore we classify this star as a DZH.

We present three new cool DC white dwarfs which have high-resolution echelle spectra from the MIKE instrument on the Magellan telescope. One of these is a newly confirmed white dwarf that lies within 1$\sigma_\varpi$ of 40\,pc, \textbf{WD\,J1935$-$0725}. In Fig.~\ref{fig:dc_mike} we demonstrate that these white dwarfs have the spectral type DC because there is no indication of an H$\alpha$ feature even with such high resolution data.

\subsection{White Dwarfs Missing from our 40 pc Spectroscopic Sample} 
\label{sec:missing}

There are five candidate white dwarfs in the catalogue of \citet{Gentile2021} that are within 40\,pc but do not have medium or high-resolution spectroscopic observations to confirm their classification. These are presented in Table~\ref{tab:WDsnospectra}. Three of these five candidates have low-resolution \gaia\ spectra available. However, the low S/N of the \gaia\ spectra mean that spectral features are not resolved and these objects cannot be confirmed as white dwarfs. Therefore medium-resolution spectroscopy with a higher S/N is still required to confirm these candidates. 

Two of these objects have a white dwarf probability factor ($P_{\rm WD}) < $ 0.75 from \citet{Gentile2021}. \textbf{WD\,J0812$-$2616} has a $P_{\rm WD}$ of 0.303, and has astrometric noise greater than 5\,mas. \textbf{WD\,J0413$-$2122} has a $P_{\rm WD}$ of 0.657. Both candidates have low \gaia\ \logg\ values. Fitting unresolved double degenerate systems as if they are single stars causes their \logg\ and mass to be underestimated, so these two candidates are likely unresolved double degenerates.

The other three candidates have $P_{\rm WD} > $ 0.75. \textbf{WD\,J1150+2404} is potentially a very red and ultra-cool white dwarf. The SDSS $g - z$ colour for \textbf{WD\,J2246$-$0609} of almost 6.0 is consistent with an isolated brown dwarf and as such it is likely a contaminant in the white dwarf sample. \textbf{WD\,J0959$-$5027} is likely to be a standard 7000\,K white dwarf, but is in the Galactic plane so is difficult to observe spectroscopically.

There are also 15 known white dwarfs within 1$\sigma_\varpi$ of 40\,pc, 12 of which have been spectroscopically confirmed. We present these white dwarfs in Table~\ref{tab:onesigma}. All white dwarfs were taken from the catalogue of \citet{Gentile2021}, except \textbf{WD\,J0548$-$7507} (see \citealt{OBrien2023} for details). We include these stars in our statistical analysis of the space density of white dwarfs, for completeness (see Section~\ref{sec:space_density}). 

There are 28 known white dwarfs within 40\,pc that are not present in the catalogue of \citet{Gentile2021}. These objects are presented in Table~\ref{tab:missingWDs}. Not all white dwarfs in this list are spectroscopically confirmed, as some were detected from radial velocity variations in main-sequence stars, rendering spectral classification difficult. Binaries will be discussed further in Section~\ref{sec:binaries}.

The star WISE\,1028$-$6327 is tentatively classified as a DAZ white dwarf in \citet{Kirkpatrick2016}, but is missing from the \citet{Gentile2021} catalogue. Its very faint absolute magnitude of $G_{\rm abs}$ = 18.5 is not consistent with a DA white dwarf and the $JHK$ flux is consistent with a M dwarf or brown dwarf. The spectrum shown in \citet{Kirkpatrick2016} is not consistent with a binary (white dwarf + M dwarf) or the \gaia\ colours. Therefore we do not include this white dwarf in our 40\,pc list.

\subsection{Spectroscopic Biases}

\begin{figure*}
    \centering
	\includegraphics[width=\textwidth]{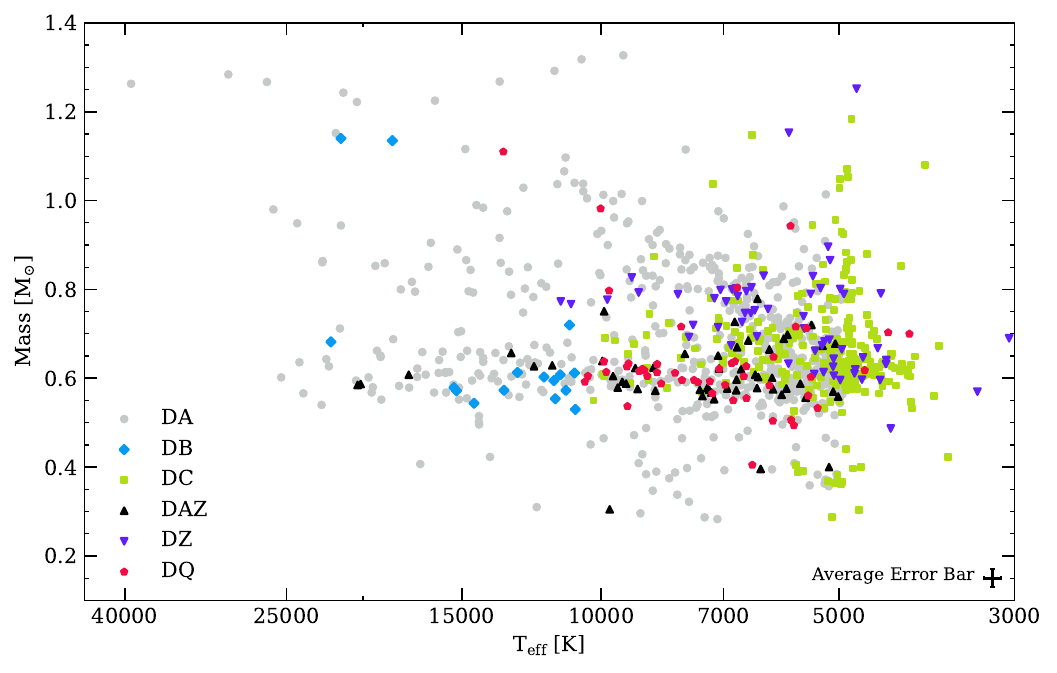}
	\caption{Corrected \gaia\ mass and \Teff\ for all 40\,pc white dwarfs, where spectral type is indicated by the shape and colour of points. White dwarfs with unreliable \gaia\ atmospheric parameters have been plotted with their parameters from Table~\ref{tab:bad_params}. The average statistical error is shown in the black point on the lower right. More complex spectral types are simplified to their most prominent features.}
    \label{fig:massteff_spt}
\end{figure*}

In Fig.~\ref{fig:massteff_spt} we show the full \gaia\ 40\,pc white dwarf sample with spectral types indicated. We include those white dwarfs with \gaia\ photometric parameters, as well as those with unreliable \gaia\ atmospheric parameters for completeness, and for which we take parameters from the literature (See Table~\ref{tab:bad_params}). For the analysis in this paper based on white dwarf mass and \Teff, we only consider white dwarfs that have reliable parameters determined from \gaia\ photometry, to keep the sample parameters homogeneous.

The 40\,pc \gaia\ white dwarf sample is spectroscopically heterogeneous $-$ spectra confirming the white dwarfs in the sample have been collected from a wide range of instruments with varying resolution and wavelength coverage. In almost all cases, spectral types are taken from optical medium-resolution ($R > $1000) spectroscopy at S/N $>$ 30. In many cases, the white dwarf has been observed multiple times at different facilities, and not all observations found in public archives are published. Therefore, it is outside of the scope of this work to list the average or best-achieved S/N, instrumental resolution and wavelength coverage for the white dwarfs in the sample. 

There are inherent issues with using a spectroscopically heterogeneous sample. Not every white dwarf has been observed with the resolution required to identify very weak signatures of metal pollution in the Ca\,\textsc{ii} H+K lines, like the kind seen in the high-resolution survey of DA white dwarfs from \citet{Zuckerman2003}. Similarly, not every white dwarf has spectropolarimetric observations or has observations at the resolution required to see faint Zeeman splitting of the spectral features, meaning the magnetic sub-sample is currently incomplete. Some DC, DB, and DZ white dwarfs may display a weak H$\alpha$ feature which would not be detected without coverage of the H$\alpha$ region \citep{OBrien2023}.


\section{Discussion}
\label{sec:disc}

\subsection{Spectral Evolution}
\label{sec:specEvolution}

\begin{figure}
    \centering
	\includegraphics[width=\columnwidth]{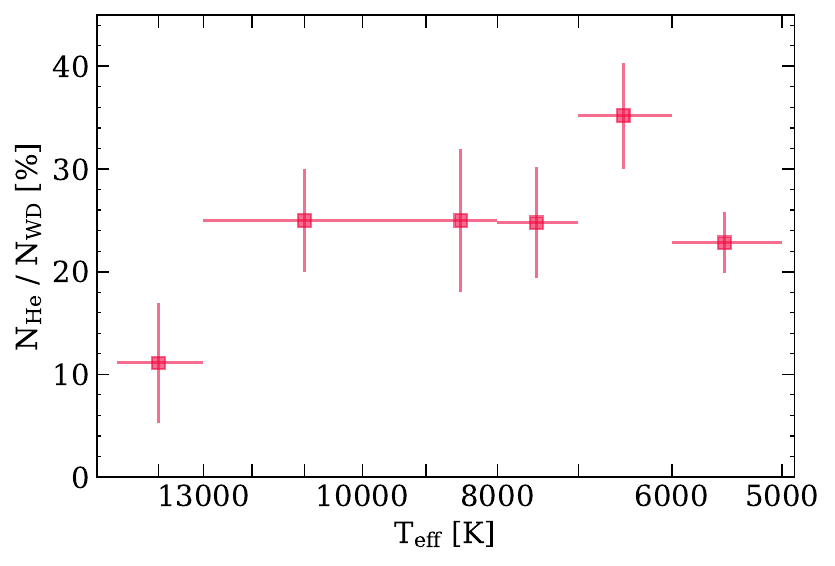}
	\caption{The fractional distribution of He-atmosphere white dwarfs compared to the full 40\,pc white dwarf sample. Horizontal error bars represent \Teff\ bins and vertical error bars show the uncertainty of the frequency of the occurrence of He-atmosphere white dwarfs within each bin.}
    \label{fig:spectral_evolution_He_teff}
\end{figure}

The atmospheric composition of a white dwarf can change with time due to physical processes including convection, atomic diffusion and accretion. In Fig.~\ref{fig:spectral_evolution_He_teff}, we study the evolution of the fraction of He-rich atmosphere white dwarfs as a function of \Teff. We note that there are few very young and hot white dwarfs in our sample, so we do not extend this study beyond 15\,000\,K. Below 5000\,K there will be no visible H$\alpha$ line in the white dwarf spectrum, so the atmospheric composition cannot be directly constrained. Therefore in Fig.~\ref{fig:spectral_evolution_He_teff} we also do not extend to \Teff\ below 5000\,K. Our observations are consistent with earlier results for the 40\,pc northern hemisphere sample \citep{Mccleery2020}. The increase in incidence between 17\,000\,K and 9000 K, which is marginal in the 40\,pc sample, has previously been attributed to convective mixing using larger samples \citep{Tremblay2008,Ourique2020,Cunningham2020,Lopez2022,Bedard2022_ii}.

Fig.~\ref{fig:spectral_evolution_He_teff} suggests an increase in the fraction of He-atmosphere white dwarfs in the range 7000\,$>$\,\Teff\,$>$\,6000\,K, although only at 1$\sigma$. This excess could be a consequence of model atmospheres with incorrect trace fractions of C and H (see Section\,\ref{sec:sample}), which in turn would result in an incorrect temperature scale. The temperature range 5000 $-$ 6000 K has been referred to as the “non-DA gap", where a decrease in the fraction of He-rich atmosphere white dwarfs was initially identified by \citet{Bergeron1997, Bergeron2001}. However, there is no clear evidence of spectral evolution in our observations at the 2$\sigma$ level in the 40\,pc sample, as shown in Fig.~\ref{fig:spectral_evolution_He_teff}.

\subsection{Mass Distributions}
\label{sec:massDistribution}

\begin{figure}
    \centering
	\includegraphics[width=\columnwidth]{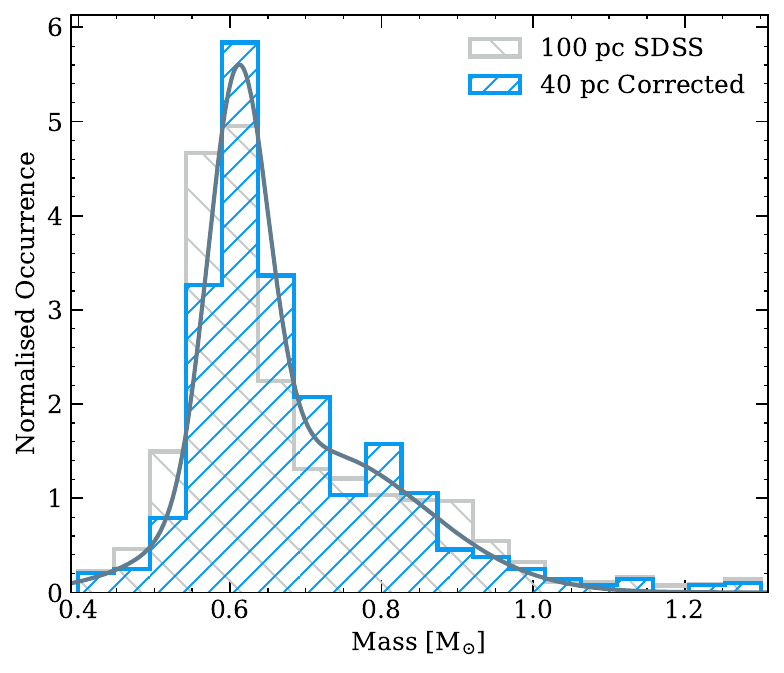}
	\caption{Mass distribution of white dwarfs in the 40\,pc sample compared to the published masses from the 100\,pc SDSS sample of \citet{Kilic2020_100pc}. We show the 40\,pc distribution with the mass correction outlined in Section~\ref{sec:corrections}. The solid grey line represents the bimodal best-fitting Gaussians to the 40\,pc mass distribution.}
    \label{fig:kilic_mass_distribution}
\end{figure}

\citet{Kilic2020_100pc} produced a volume-like sample of DA white dwarfs within 100\,pc in the SDSS footprint, with \Teff\ $>$ 6000\,K. In Fig.~\ref{fig:kilic_mass_distribution} we show that the peak of the mass distribution of our 40\,pc sample with corrected photometric \gaia\ masses is in a similar position to that of the 100\,pc SDSS sample, where photometric masses were derived from SDSS $u$, Pan-STARRS $grizy$ and \gaia\ DR2 parallaxes \citep{Kilic2020_100pc}.

\citet{Kilic2020_100pc} observe a peak in the 100\,pc DA mass distribution at 0.59\,\Msun, and a `shoulder' at 0.7\,\Msun\ $-$ 0.9\,\Msun. In Fig.~\ref{fig:kilic_mass_distribution}, we similarly fit two Gaussian curves, one to the main peak and one to the prominent shoulder of the 40\,pc mass distribution. The main peak in our distribution sits at 0.61 \Msun\ with a standard deviation of 0.07 \Msun. In Fig.~\ref{fig:kilic_mass_distribution} there is a notable secondary intermediate-mass peak, or shoulder, the cause of which remains elusive. \cite{Kilic2020_100pc} have demonstrated, based on binary population synthesis models \citep{Temmink2020}, that the single white dwarfs formed from mergers cannot be the dominant explanation for a shoulder in the white dwarf mass distribution. 

\citet{Kilic2020_100pc} have suggested that the shoulder could be attributed to white dwarf core crystallisation, although our sample is volume complete and crystallisation cooling delays should not influence the mass distribution as we plot all white dwarfs at all \Teff\ values. Furthermore, samples of warm, non-crystallised white dwarfs show this shoulder \citep{Tremblay2016, Sahu2023}. Fig.~\ref{fig:massteff_spt} also displays a distinct branch of 0.8--0.9\,\Msun\ white dwarfs at 10\,000--25\,000\,K, separated from the main distribution at 0.6\,\Msun, and the crystallisation branch at higher masses. Another explanation given by \cite{Tremblay2016,Tremblay2019} and \cite{El-Badry2018} is that the shoulder or secondary peak is caused by a flattening of the initial-final mass relation (IFMR) at initial masses $3.5\leq\rm{M}/\Msun\leq 4.5$ \citep{Cummings2018}, leading to an accumulation of white dwarf at masses $\sim$0.8\,\Msun. This is possibly linked to the the onset of the second dredge-up in asymptotic giant branch stars \citep{marigo2007,cummings15}.

We have artificially corrected the photometric \gaia\ masses for white dwarfs with \Teff\ $<$ 6000\,K (see Section~\ref{sec:corrections}). The mean \gaia\ mass for white dwarfs with \Teff\ $>$ 6000\,K is 0.69 $\pm$ 0.01\,\Msun. The mean mass of the 40\,pc northern sample from \citet{Mccleery2020} for \Teff\ $>$ 5000\,K is slightly lower, 0.66\,\Msun, however in the current full 40\,pc sample the masses were corrected for \Teff\ $<$ 6000\,K so these means cannot be directly compared.

There are 33 white dwarfs in the sample with masses $>$ 1\,\Msun\ (three\,per\,cent of the sample). Just over a third of these are magnetic, the impact of which is discussed in Section~\ref{sec:magnetic}. Of these, 79\,per\,cent have H-dominated atmospheres, which is comparable to the 72\,per\,cent of the full sample with \Teff\ above 5000\,K which have H-dominated atmospheres.

\subsection{Multiple-star systems containing white dwarfs}
\label{sec:binaries}

Considering all known and candidate white dwarfs within 40\,pc (see Section~\ref{sec:space_density} for details), and considering all double degenerate candidates, we find that there are 209 multiple-star systems within 40\,pc containing at least one white dwarf. Based on binary population synthesis models when considering a constant stellar formation history as was found for the 40\,pc sample by \citet{Cukanovaite2023}, we predict using \citet{Toonen2017} models that there should be 318--458 binary systems containing a white dwarf within 40\,pc. This discrepancy partially originates from the lack of wide double white dwarfs which will be discussed in Section~\ref{sec:wideBinaries}. 

\subsubsection{Wide Binaries}
\label{sec:wideBinaries}

We differentiate between wide and unresolved binaries based on whether they are resolved as separate sources in \gaia\ DR3 or not, where the \gaia\ on-sky resolution is 0.4\,arcsec. We also discuss triple systems, for which at least one component is resolved in \gaia. In this section, we use WD to denote a white dwarf, MS to denote a main-sequence star, and BD to denote a brown dwarf. All wide binaries and higher order systems with at least one white dwarf companion within 40\,pc are presented in Table~\ref{tab:widebinaries_online}.

\begin{figure}
    \centering
	\includegraphics[width=\columnwidth]{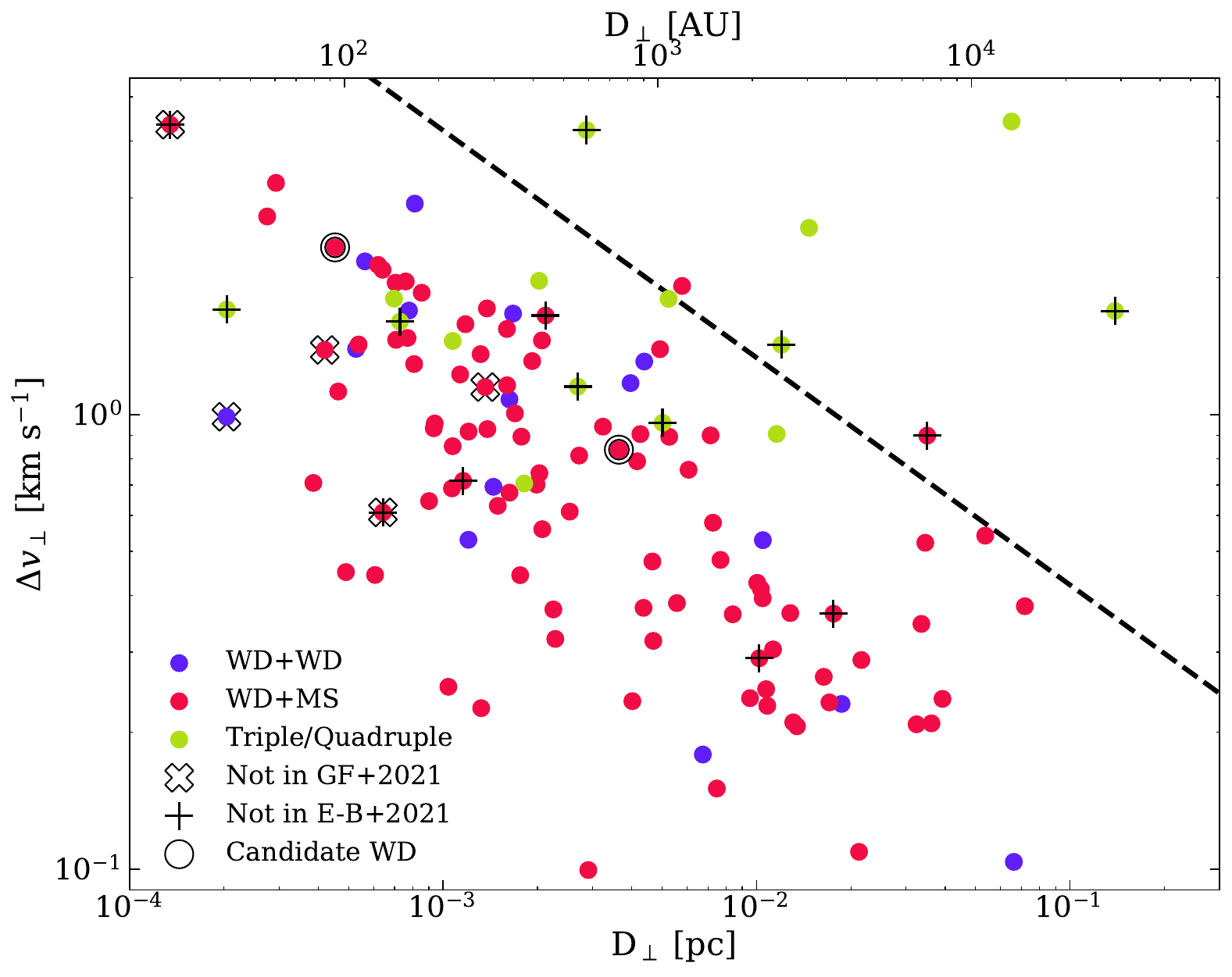}
	\caption{Tangential velocity differences as a function of projected separation for \gaia\ DR3 sources within 1\,pc of the white dwarfs in the 40\,pc sample. The black dashed line indicates the maximum tangential velocity difference for a binary with 1.4\,\Msun+2.5\,\Msun\ stars on a circular orbit. WD+WD systems are shown in purple, WD+MS systems in red, and triple or quadruple systems in green.}
    \label{fig:widebinaries}
\end{figure}

To search for resolved \gaia\ common proper motion companions to the white dwarfs in our sample, we used the same strategy as described in \citet{Hollands2018_Gaia} and \citet{Mccleery2020}. In short, we performed a cone-search for each white dwarf with a \gaia\ parallax greater than 25\,mas, scaled by distance, for a projected separation of 1\,pc and radial distance within 1\,pc of the white dwarf. The tangential velocity difference is obtained using the difference in proper motion for the two stars as given in \gaia\ DR3. We recover 121 wide binaries and triples from this search, which are displayed in Fig.~\ref{fig:widebinaries}. We also recover 33 contaminant pairs in our search, which we remove. These sources either have large parallax errors, are in crowded regions of the Galaxy, or have unphysical separations in velocity-separation space if they are a binary pair. We show a dashed line on Fig.~\ref{fig:widebinaries} which indicates the maximum difference in tangential velocity for a WD+MS binary system with component masses 1.4\,\Msun\ and 2.5\,\Msun\ respectively \citep{Torres2022}.

A few systems that are above the dashed line in Fig.~\ref{fig:widebinaries} are known to be genuine wide systems, where their inconsistent separation and velocity difference is caused by higher-order multiplicity. \textbf{WD\,J2101$-$4906} has a main-sequence companion that is itself an unresolved binary \citep{Hollands2018_Gaia}, and \textbf{WD\,J1702$-$5314} also has an unresolved binary as a companion. \textbf{WD\,J2004+0109} is a wide companion to a spectroscopic triple main-sequence system \citep{Venner2023}. \textbf{WD\,J0103+0504} is an unresolved double white dwarf with a double main-sequence binary companion. We inspected systems lying near the dashed line in Fig.~\ref{fig:widebinaries}, and kept three such systems in our final wide binary catalogue for completeness.

There are six WD(+MS)+BD systems where the brown dwarf is not in \gaia\ DR3 \citep{Leggett2015, Mace2018, Meisner2020, Zhang2020, Gonzales2022}. There are three known quadruple systems: one of which comprises \textbf{WD\,J0103+0504} which is a double degenerate, plus K-type stars HD\,6101 A+B \citep{Mccleery2020}; the other which has two resolved white dwarfs from \citet{Limoges2015} that are not in \citet{Gentile2021} (\textbf{WD\,0727+482A} and \textbf{B}) with an unresolved pair of M-dwarfs as a wide companion. \textbf{WD\,J2004+0109} is also part of a quadruple system as mentioned above. We add these to our list, alongside other known systems missing from \gaia, and therefore obtain 132 wide binaries, triples, and quadruples in total within 40\,pc. These systems are classified as follows: 97 WD+MS, 15 WD+WD, 9 WD+MS+MS, 1 WD+WD+MS, 1 WD+WD+WD, 2 WD+WD+MS+MS, 1 WD+MS+MS+MS, 5 WD+BD, 1 WD+MS+BD. For 8 of these systems, the white dwarf is missing from the catalogue of \citet{Gentile2021}, because the bright main-sequence companion affects the white dwarf colours or astrometry. These systems are shown as cross symbols in Fig.~\ref{fig:widebinaries}, where proper motions are available.

\citet{El-Badry2021} searched for wide binaries in \gaia\ DR3, but intentionally removed triple and quadruple systems. They do not recover seven WD+MS systems that we find in our cone search, for which \gaia\ proper motions were available for both stars. These are shown as plus symbols in Fig.~\ref{fig:widebinaries}. Five of these missing systems have a white dwarf that is in \citet{Gentile2021}. For one of these systems, the main-sequence star has a candidate close brown dwarf companion, which would make this system a triple \citep{Diaz2012}. Another missing system, $\varepsilon$ Reticuli A+B, also has a close gas giant planet orbiting the main-sequence component \citep{Butler2001}. 

\citet{Toonen2017} predict 169--228 resolved WD+MS systems within 40\,pc, compared to the 97 that we observe, and 119--167 resolved WD+WD systems compared to the 15 that we observe. The notable lack of wide WD+WD systems in volume-limited samples compared to predictions from binary population models is discussed in \citet{Toonen2017,El-Badry2018,Mccleery2020}.

We find two metal-polluted white dwarfs out of 90 (2\,per\,cent) that are in wide binaries with projected separations between 120 and 2500\,au, and two polluted white dwarfs out of 22 (10\,per\,cent) in wide binaries with projected separations greater than 2500\,au. The overall fraction of white dwarfs displaying signs of pollution within 40\,pc is 11\,per\,cent. Therefore our findings are in line with predictions made by \citet{Zuckerman2014}, indicating that a close secondary star can suppress the formation or retention of a planetary system (see also \citet{Wilson2019} for further discussion).

\subsubsection{Unresolved Binaries}
\label{sec:unresolvedBinaries}

\begin{figure}
    \centering
	\includegraphics[width=\columnwidth]{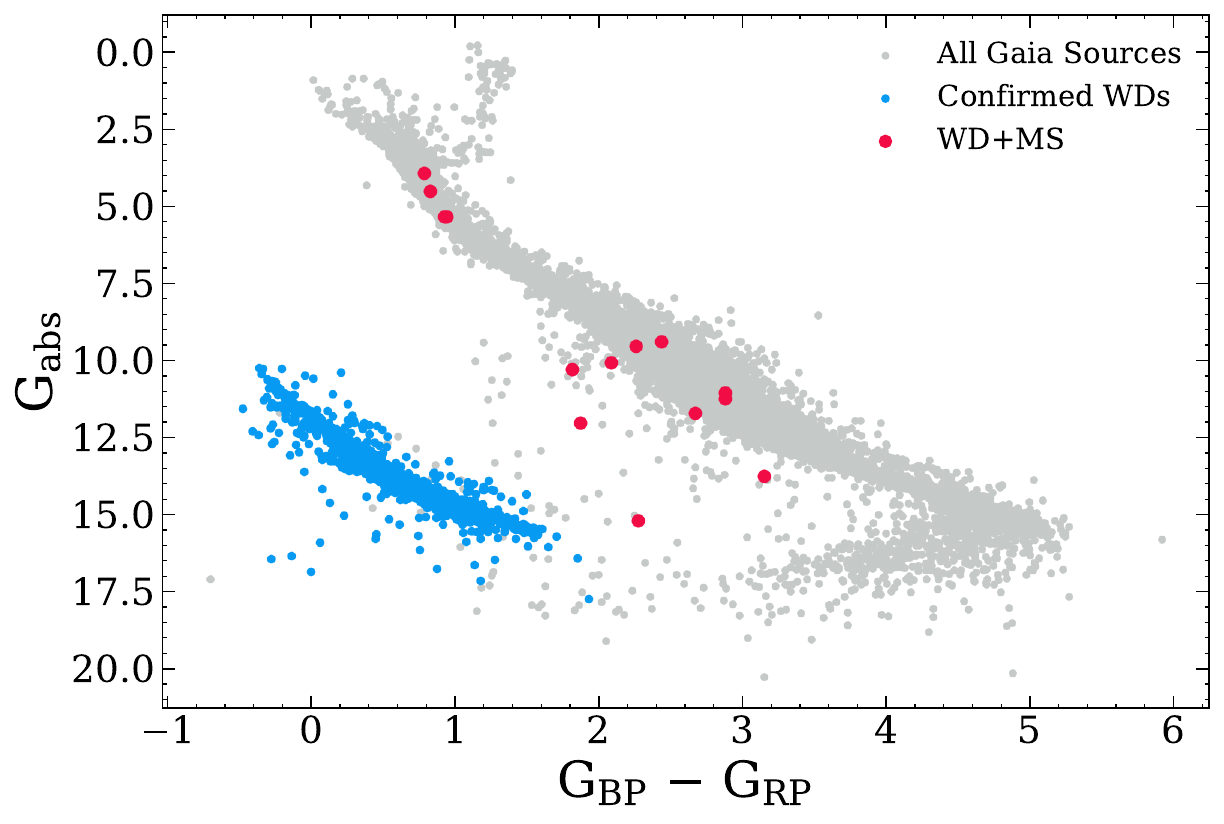}
	\caption{A \gaia\ HR diagram showing all spectroscopically confirmed white dwarfs within 40\,pc as blue points, alongside all other \gaia\ sources in the same volume which have \texttt{parallax\_over\_error} $>$ 1 and \texttt{astrometric\_excess\_noise} $<$ 1.5 from \gaia\ DR3 as grey points. White dwarfs in unresolved binaries with main-sequence companions that are not in the \citet{Gentile2021} catalogue are shown as red points.}
    \label{fig:fullGaiaHR}
\end{figure}

There are five unresolved WD+MS systems in the \gaia\ 40\,pc sample that have white dwarfs in the \citet{Gentile2021} catalogue. There are also 19 unresolved WD+MS systems that are not in our main sample (see Table~\ref{tab:missingWDs}), and one extra unresolved WD+MS within 1$\sigma_\varpi$ of 40\,pc. Many of these systems that are missing from \citet{Gentile2021} consist of cool white dwarfs with main-sequence companions, such that their \gaia\ photometry places them on or close to the main sequence (see Fig.\,\ref{fig:fullGaiaHR}). These systems have been serendipitously detected in the literature due to an ultraviolet excess from the white dwarf, photometric variability or radial velocity measurements. Regulus A+B, Procyon A+B and HD\,149499 A+B are known WD+MS binaries, but the white dwarf is not in \gaia\ at all. Our sample is likely to be incomplete until systematic spectroscopic, photometric and astrometric variability searches are performed for all $\approx$ 20\,000 main-sequence stars within 40\,pc. 

The \citet{Cummings2018} initial-final mass relation breaks down for single-star evolution at 0.53\Msun. White dwarfs below this mass cannot be produced in isolation within the lifetime of the Milky Way \citep{Marsh1995}. If the photometry of unresolved WD+WD (double-degenerate) systems is fitted as if they were single stars, their masses are underestimated (and radii overestimated) and they lie below the median mass sequence. There are 54 white dwarfs within 40\,pc that have corrected \gaia\ masses less than 0.49\,\Msun, which is 2\,$\sigma$ below the mass limit for single-star evolution assuming an average photometric mass error of 0.02\,\Msun, and these stars are therefore highly likely to be double degenerate systems. All candidate and confirmed unresolved WD+WD systems are flagged in the comment column of Table~\ref{tab:all_online}. 

Fourteen of these systems within 40\,pc are confirmed as unresolved WD+WD binaries from radial velocity measurements \citep{Napiwotzki2020, Kilic2020_WDWD, Kilic2021}. Two of these binaries also have a third wide companion. 

If one of the white dwarfs in a double degenerate system is a low-mass white dwarf, formed through mass transfer in the binary evolution process, only the bright low-mass component may be detectable for both the spectrum and photometry of the system. For a featureless double degenerate spectrum, there is no way of determining the individual white dwarf masses in the binary. However, this is possible for DA spectral types. The mass for \textbf{WD\,J0946+4354} determined from spectroscopy is 0.45\,\Msun\ \citep{Limoges2015} compared to a photometric mass of 0.42\,\Msun, indicating that this system contains a genuine low-mass white dwarf. Similarly, \textbf{WD\,J0841$-$3256} has a photometric mass of 0.47\,\Msun\ and a spectroscopic mass of 0.45\,\Msun\ \citep{Bedard2017}. Due to the heterogeneous nature of the 40\,pc sample, not all white dwarfs have parameters determined from spectroscopy, and therefore all WD+WD candidates should be followed up for further study to search for more low-mass white dwarfs.

Eleven of the candidate double degenerate systems additionally have a \gaia\ renormalised unit weight error (RUWE) value above 1.4, indicating poor quality astrometric solutions and a high probability of binarity. Six white dwarfs within 40\,pc have non-single-star astrometric periods from \gaia\ ranging from 77 -- 249 days, and these are shown in Table~\ref{tab:gaia_astrometric}. Five of these six are also over-luminous double degenerate candidates.

\begin{table}
	\centering
        \caption{Double white dwarf binaries with \gaia\ astrometric periods.}
        \label{tab:gaia_astrometric}
        \begin{tabular}{ll}
                \hline
                WD\,J Name & \gaia\ Orbital Period [days] \\
                \hline
                023117.04+285939.88 & 103.89 (0.08) \\
                092943.17$-$173250.68 & 238.0 (0.3) \\
                142054.81$-$090508.76 & 87.53 (0.06) \\
                200654.88+614310.27 & 77.1 (0.1) \\
                211345.93+262133.27 & 219.7 (0.2)\\
                232519.87+140339.61 & 249 (1) \\
                \hline
        \end{tabular}\\
\end{table} 

\textbf{WD\,J0948+2421}, as mentioned in \citet{Mccleery2020}, has a larger than average \gaia\ mass of 0.80 $\pm$ 0.01\,\Msun, but is a known DA+DAH system comprised of two more massive white dwarfs \citep{Liebert1993}. Similarly, \textbf{WD\,J0138$-$1954} is a double-lined DA+DA binary from \citet{Napiwotzki2020} with a large \gaia\ mass of 0.93 $\pm$ 0.01\,\Msun, with the combined low luminosity suggesting a pair of ultra-massive white dwarfs.

Based on binary population synthesis models with a constant stellar formation history within 40\,pc, \citet{Toonen2017} predict 6--12 unresolved WD+MS systems compared to our 25, and 24--51 unresolved WD+WD systems compared to our upper limit of 54. Furthermore, extrapolating from the \citet{Hollands2018_Gaia} 20\,pc white dwarf sample predicts the number of unresolved binaries expected in 40\,pc relatively well: 16 unresolved WD+MS and 56 unresolved WD+WD systems. The models appear to under-predict the numbers of unresolved WD+MS -- however the numbers of these systems are not well constrained by observations as they are difficult to detect. 

\subsection{Space Density}
\label{sec:space_density}

\begin{figure}
    \centering
	\includegraphics[width=\columnwidth]{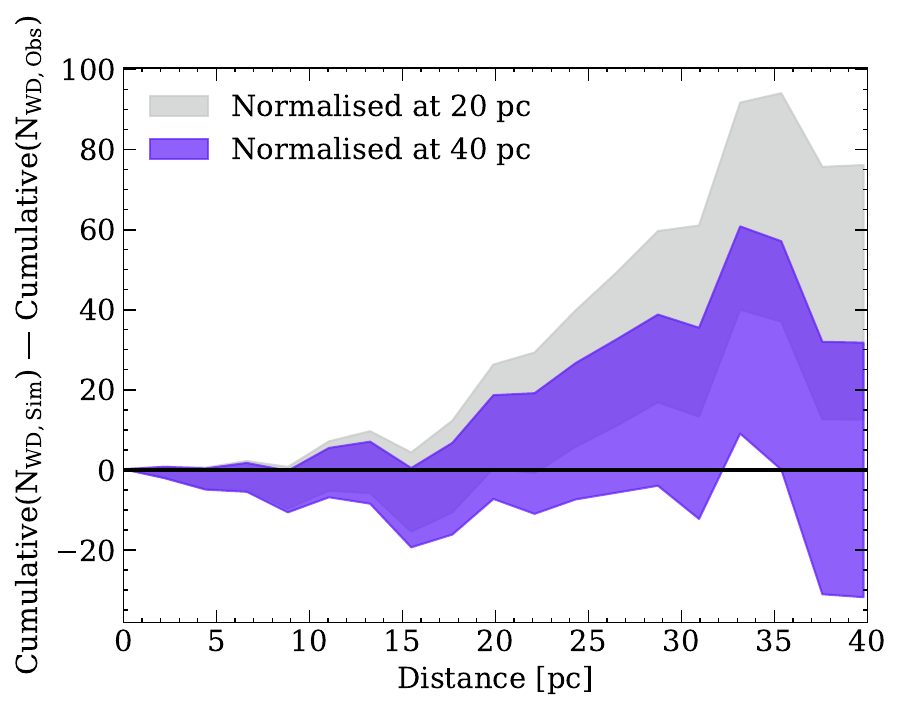}
	\caption{The space density of white dwarfs according to our Galactic simulation as function of the distance to the Sun, compared to the observed space density from the 40\,pc sample. The simulation has been normalised at both 20\,pc and 40\,pc separately. The filled regions represent an error of 1$\sigma$ on the numbers of white dwarfs. $\mathrm{N_{WD,Sim}}$ is the number of simulated white dwarfs, and $\mathrm{N_{WD,Obs}}$ is the number of observed white dwarfs.}
    \label{fig:simulation}
\end{figure}

In Fig.~\ref{fig:simulation} we show the results of a Galactic simulation of a single white dwarf population carried out with the same initial conditions as that described in \citet{Cukanovaite2023}, with a million white dwarfs simulated within 40\,pc. The vertical position of the Sun is set to 20\,pc above the Galactic plane in the simulation, while the vertical scale height of the Galactic disk varies according to the observed white dwarf velocity dispersion. However, the absolute value of the vertical scale height is fixed at 75\,pc for 1\,Gyr because it is difficult to constrain this quantity directly from faint white dwarfs. 

In the simulation, white dwarfs are only formed with \gaia\ masses above 0.54\,\Msun, as in \citet{Cukanovaite2023}. Therefore, we remove white dwarfs below this mass in our data when comparing to the simulation, resulting in 1010 white dwarfs within 40\,pc including those not in the \citet{Gentile2021} catalogue. We normalise the simulation both at 20\,pc and separately at 40\,pc. The simulation over-predicts the number of white dwarfs we would expect at 30--40\,pc by up to 2$\sigma$. This discrepancy could be an indication that the assumed Galactic scale height in the simulation is incorrect, that the assumed vertical position of the Sun above the Galactic plane is incorrect, or that we have $\approx$\,50 missing white dwarfs, possibly hidden in close binaries with bright main-sequence companions.

\citet{Gentile2021} used the \textit{Gaia} DR3 catalogue of white dwarfs within 20\,pc to infer a local space density of 4.47 $\pm$ \num{0.37e-3} pc$^{-3}$ including those missing from \textit{Gaia}. Unresolved double degenerates are counted as one system in the calculation of space density \citep[see][]{Hollands2018_Gaia}. Extrapolating by volume to 40\,pc, we would expect 1198 $\pm$ 99 white dwarfs. Considering all confirmed DR3 white dwarfs (1076), confirmed white dwarfs within 1$\sigma_\varpi$ of 40\,pc (15), remaining DR3 candidates (7), and known white dwarfs missing from \citet{Gentile2021} (28), we count 1124 possible white dwarfs within 40\,pc, within 1$\sigma$ of the extrapolated 20\,pc space density.

\subsection{Multi-Wavelength Analysis}

\begin{figure*}
	\includegraphics[viewport= 1 20 520 720, scale=0.85]{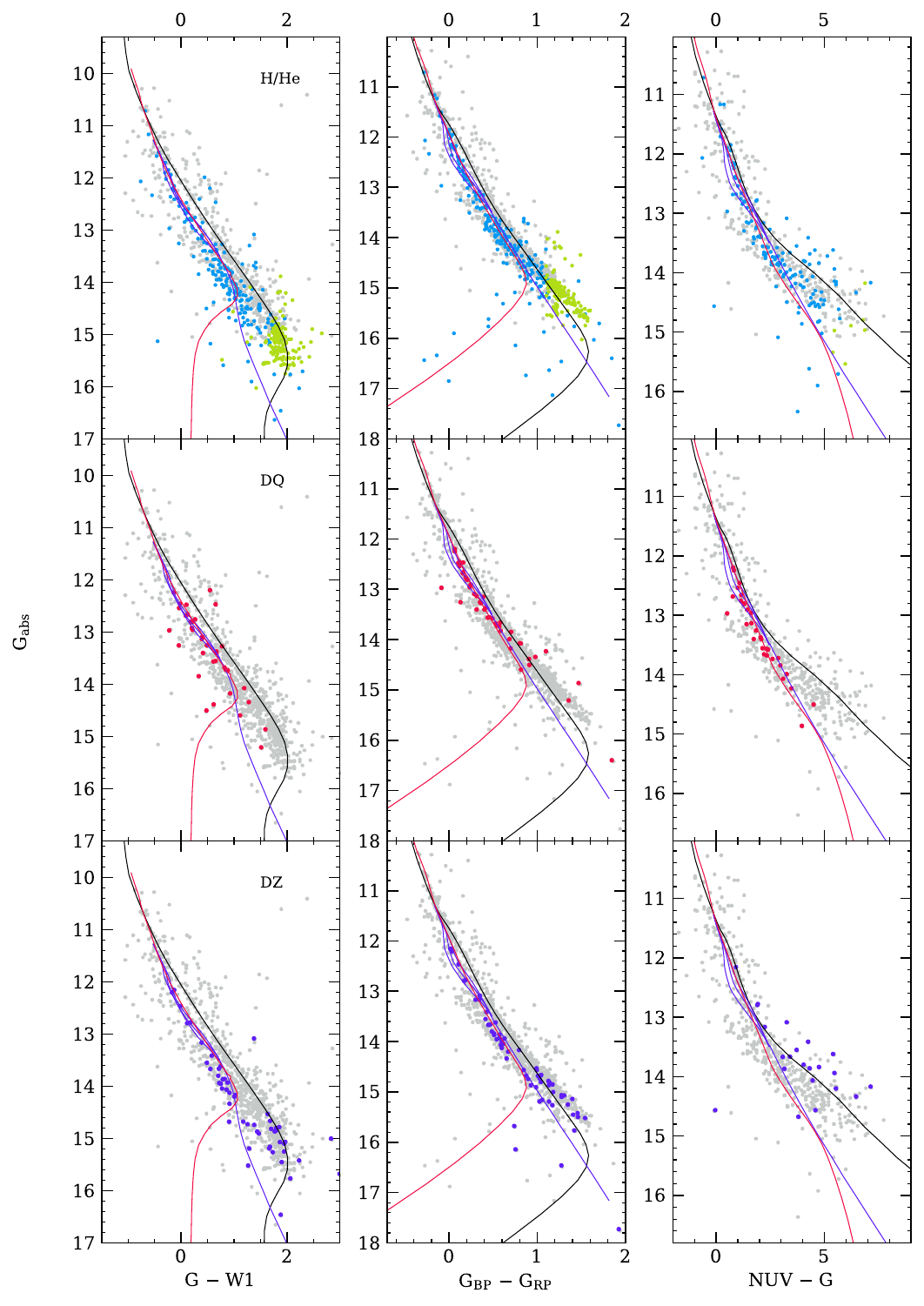}
	\vspace*{10mm}
	\caption{HR diagrams for the 40\,pc white dwarf sample. Top: H-atmosphere white dwarfs are in grey, He-atmosphere white dwarfs are in blue, and cool DC white dwarfs with unconstrained composition are in green. Middle: DQ white dwarfs are in red and the rest of the sample is in grey. Bottom: DZ white dwarfs are in purple and the rest of the sample is in grey. In all panels, The black lines indicate pure-H cooling tracks, the red lines indicate mixed H/He = $10^{-5}$, and the purple lines indicate mixed C/He cooling tracks with varying initial C mass fractions in the envelope of the PG 1159 (0.2, 0.4 and 0.6; see \citealt{Blouin2023}). In all cases, tracks are for a 0.6\,\Msun\ white dwarf.}
        \label{fig:ninepanels}
\end{figure*}

In Fig.~\ref{fig:ninepanels} we use \textit{WISE} \citep{WISE2010}, \gaia, and \textit{GALEX} \citep{GALEX2005} photometry to study the positions of white dwarfs with H and He atmospheres (top panels) on HR diagrams. We also separate out the DQ and DZ spectral types (middle and bottom panels). We apply a linearity correction to the \textit{GALEX} near-UV photometry, provided by \citet{Wall2019}. Cool DC white dwarfs are shown in the top panels of Fig.~\ref{fig:ninepanels} as green points; they are assumed to have pure-H atmospheres for the purpose of deriving atmospheric parameters but a lack of spectral features means that their composition cannot be constrained. 

There is a bifurcation observed between H and He atmosphere white dwarfs in the \gaia\ HR diagram \citep{Gaia2018,Gentile2019,Bergeron2019,Blouin2023,Camisassa2023}, commonly described as DA white dwarfs following the A-branch, and He-rich atmospheres following the B-branch \citep{Gaia2018_HR}. It is clear from Fig.~\ref{fig:ninepanels} that DB, DC, DQ, and DZ white dwarfs indeed follow the B branch. In this work and previous papers in the 40\,pc series, we use He-rich models with additional H, with a composition of $\log({\rm H/He})$ $= -$5 \citep{Mccleery2020,Gentile2021,OBrien2023}. In Fig.~\ref{fig:ninepanels}, the white dwarfs with He-dominated atmospheres (DB, DC, DQ and DZ) follow this model sequence closely, but deviate from it as they cool (e.g. $G_{\rm BP}-G_{\rm RP} < 0.8$). 
We note that the bifurcation and its agreement with models are very similar at optical and IR wavelengths, which is expected since He$^{-}$ free-free opacity, which is sensitive to free electrons from trace hydrogen or carbon, dominates at both wavelengths. At cool temperatures, CIA opacity from hydrogen sets in, which explains why mixed H/He models turn bluer. 

As discussed in Section~\ref{sec:sample}, He-rich models with trace C below the optical detection limit better reproduce the \gaia\ HR diagram bifurcation \citep{Blouin2023,Blouin2023b,Camisassa2023}. In Fig.~\ref{fig:ninepanels} we show mixed C/He cooling tracks with three different initial C mass fractions in the envelope of the PG 1159 star progenitor to the white dwarf: 0.2, 0.4 and 0.6 \citep{Blouin2023}. The C abundance is not fixed in these models, but instead follows the evolutionary predictions of \citet{Bedard2022_iii}. For \Teff\ $<$ 7000\,K, the C abundance is so low that pure-He models are appropriate. We use pure-He models from \citet{Blouin2018} at these cool \Teff\ values. These C/He (pure-He) cooling tracks provide a better fit to the \gaia\ HR diagram than H/He tracks for faint white dwarfs ($G_{\rm abs} \gtrsim 15$). However both cooling tracks fit poorly in the IR for the same regime, suggesting that additional physical issues need to be solved in the models \citep{Saumon2022} before spectral evolution of trace H and C can be studied in low temperature ($<$ 7000\,K) He-atmosphere white dwarfs.

In the UV, cool DA white dwarfs lie below the pure-H sequence for $G_{\rm abs} \gtrsim 13.5$ whereas in the optical they lie above the pure-H sequence for $G_{\rm abs} \gtrsim 15$, corresponding to the \gaia\ low-mass problem at low temperatures ($<$ 6000\,K) discussed in Section~\ref{sec:corrections}. Re-scaling the Ly$\alpha$ opacity in the models improves the fit to the optical HR diagram but worsens the fit to the UV HR diagram, indicating that a simple multiplication factor of the Ly$\alpha$ opacity is an incomplete solution to the opacity problem of cool white dwarfs.

DZ white dwarfs appear redder than expected in the bottom right panel of Fig.~\ref{fig:ninepanels}, the \gaia$-$\textit{GALEX} HR diagram, when compared to H/He cooling tracks. UV flux suppression in DZ white dwarfs compared to He-atmosphere white dwarfs is expected because of the large number of UV metal absorption lines \citep{Wolff2002}. Flux is therefore emitted at redder wavelengths to produce the same overall flux corresponding to the white dwarf \Teff.
 
\subsection{Magnetic White Dwarfs}
\label{sec:magnetic}

\begin{figure}
    \centering
	\includegraphics[width=\columnwidth]{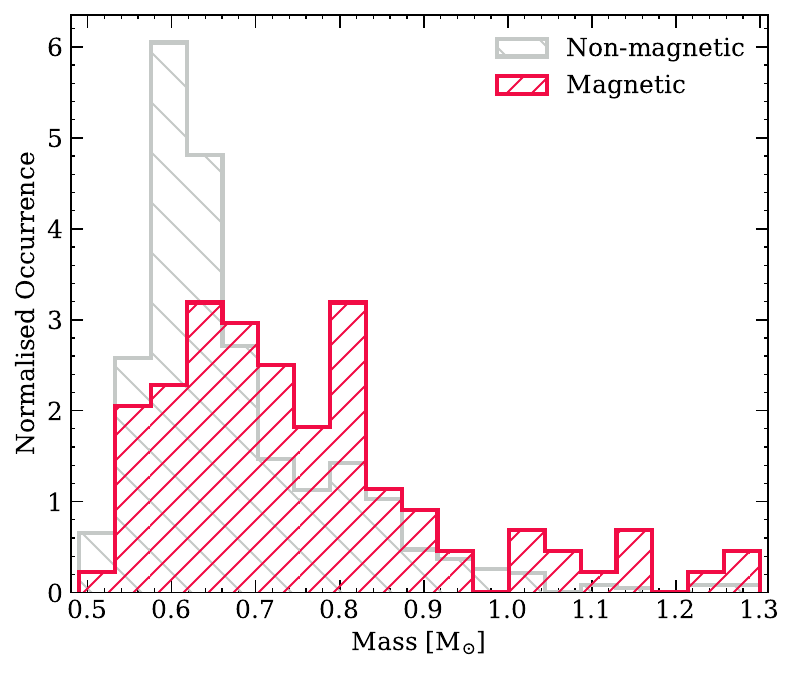}
	\caption{The mass distribution of magnetic (red) and non-magnetic (grey) white dwarfs within 40\,pc.}
    \label{fig:magnetichist_mass}
\end{figure}

\begin{figure}
    \centering
	\includegraphics[width=\columnwidth]{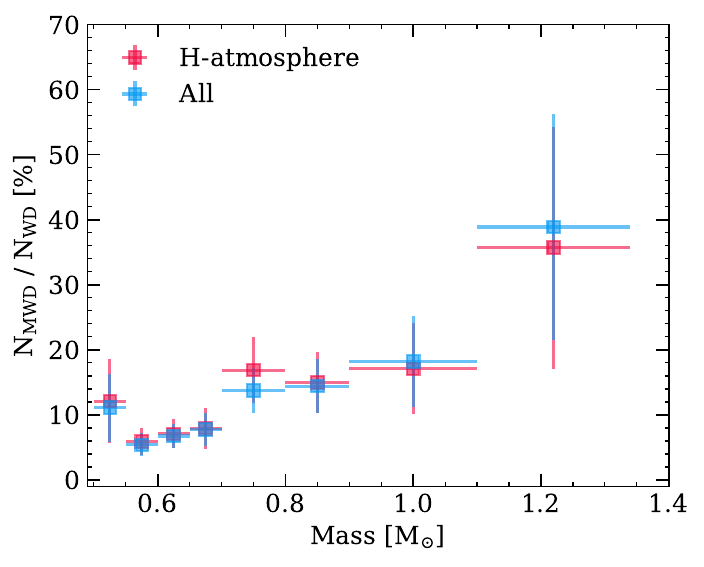}
	\caption{Frequency of magnetic white dwarfs as a function of mass for white dwarfs within 40\,pc and \Teff\ $>$ 5000\,K, for H-dominated atmospheres (red) and for all spectral types (blue). Horizontal error bars represent mass bins and vertical error bars show the uncertainty of the frequency of the occurrence of magnetic fields within each bin.}
    \label{fig:magneticfreq_mass}
\end{figure}

\begin{figure}
    \centering
	\includegraphics[width=\columnwidth]{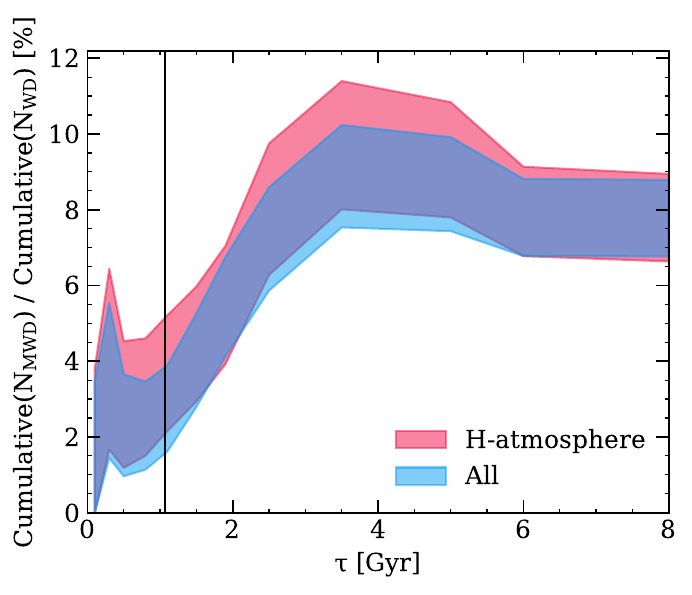}
	\caption{Fractional cumulative frequency of magnetic white dwarfs as a function of cooling age for corrected masses between 0.49\,\Msun\ and 0.8\,\Msun, for all white dwarfs within 40\,pc that have H-dominated atmospheres (red) and all white dwarfs within 40\,pc (blue). $\tau$ indicates the cooling age of the white dwarf. Cooling ages were derived using updated \textsc{STELUM} evolutionary models \citep{Bedard2022_ii,Elms2023}, and the filled regions cover 1$\sigma$ uncertainty. The black line indicates the cooling age of the onset of crystallisation for core oxygen mass fraction X(O) $=$ 0.8 for a white dwarf with a mass of 0.8\,\Msun.}
    \label{fig:magneticfreq_age}
\end{figure}

A fraction of local white dwarfs have been observed to host magnetic fields, ranging in strength from 30\,kG to several hundreds of MG \citep{Kawka2012,Bagnulo2021,Bagnulo2022,Hardy2023,Hardy2023b}; and their origin is not well constrained. These fields are usually detected through direct observations of the Zeeman splitting of white dwarf spectral features, although dedicated searches using spectropolarimetry have also discovered many of these magnetic stars (e.g. \citealt{Bagnulo2018,Bagnulo2021}). Current ideas to explain magnetic fields in isolated white dwarfs include: the field was generated by a dynamo in the core of the white dwarf during the crystallisation process or the merger of two white dwarfs, the field has been generated by a dynamo in the core of the giant or main-sequence progenitors (possibly with binary/planet interaction), or a fossil field has persisted since stellar formation \citep{Briggs2015,Briggs2018,Cantiello2016,Ferrario2020,Schreiber2021,Bagnulo2021,Ginzburg2022,Bagnulo2022}. 

With the 40\,pc northern sample, \citet{Mccleery2020} demonstrate using UV and IR photometry that the \gaia\ temperatures and masses are similarly accurate for cool (\Teff\ $\lesssim$ 12\,000\,K) magnetic white dwarfs as they are for non-magnetic white dwarfs. \citet{Hardy2023} find a similar result for cool magnetic white dwarfs. This is also supported by the observation that most of the massive DAHs lie on the \gaia\ crystallisation branch (Q-branch;  that extends from the upper left corner to about 0.8\,\Msun\ and \Teff\ = 7000 K in Fig.~\ref{fig:massteff_spt}). This would be an unlikely coincidence if the \gaia-derived atmospheric parameters were inaccurate.

\citet{Mccleery2020} observed in the northern 40\,pc sample that magnetic white dwarfs have, on average, a higher mass than non-magnetic white dwarfs. In Fig.~\ref{fig:magnetichist_mass}, we see that non-magnetic white dwarfs have a narrow peak around the canonical mass of 0.6\,\Msun, whereas magnetic white dwarfs have a less prominent peak and a larger dispersion, leading to a larger average mass of 0.75\,\Msun. We also show the resulting magnetic to non-magnetic ratio as a function of mass in Fig.~\ref{fig:magneticfreq_mass}. Magnetism is much easier to detect in pure-H DA white dwarfs as they have visible spectral features down to \Teff\ $\approx$ 5000\,K. We therefore focus on DA white dwarfs above this temperature in the following discussion, but note that the same trends are observed in the full sample. For this analysis, only white dwarfs with corrected \gaia\ masses greater than 0.49\,\Msun\ are considered, to remove contamination from double degenerates (see Section~\ref{sec:unresolvedBinaries} for details). 

\citet{Bagnulo2022} infer that there are two populations of magnetic white dwarfs with very different typical masses. They searched for magnetism using spectropolarimetry for all white dwarfs within 20\,pc and white dwarfs younger than 0.6\,Gyr within 40\,pc. In very massive white dwarfs, they find magnetic fields to be common at short cooling ages. In lower-mass white dwarfs, magnetic fields are rare, but their incidence grows with cooling age. These two populations of magnetic white dwarfs are hinted in Figs.~\ref{fig:magnetichist_mass} and~\ref{fig:magneticfreq_mass}. There is a peak in the fraction of magnetic white dwarfs at around 0.8\,\Msun, but there is also a tentative peak at higher masses ($>$1.1\,\Msun). However, we note that not enough of these massive magnetic white dwarfs have been observed out to 40\,pc to constrain the significance of this second peak. 

In Fig.~\ref{fig:magneticfreq_age}, we see that for white dwarfs with masses $<$\,0.8\,\Msun, magnetic fields emerge 1--3\,Gyr after the star becomes a white dwarf, similar to the observations of \citet{Bagnulo2022}. This result was missed by \citet{Mccleery2020} due to their smaller sample size and lack of mass cutoff. This age is around the time at which white dwarfs begin to crystallise.

We determine the age of the onset of crystallisation for all our magnetic white dwarfs with masses 0.5\,\Msun\,$<$\,\textit{M}\,$<$\,0.8\,\Msun. We test the effect of changing the assumed core oxygen mass fraction, X(O), on the predicted onset of crystallisation. For X(O) $=$ 0.60, we find that 55 ($\pm$ 4) out of 69 systems have already begun crystallising, while for X(O) $=$ 0.8 we find that 61 ($\pm$ 2) out of 69 have begun crystallising, where errors indicate those that are within 3\,$\sigma$ of the age of the onset of crystallisation. The earliest possible age at which crystallisation could begin according to our models, with X(O) $=$ 0.80 and 0.8\,\Msun, is shown by the black line in in Fig.~\ref{fig:magneticfreq_age}, which was calculated using updated crystallisation models from the \textsc{STELUM} code \citep{Bedard2022_ii,Elms2023}. It is clear that some magnetic systems lie to the left of that line. X(O) $=$ 0.60 is a more standard abundance based on pre-white dwarf evolution models, however even with more lenient conditions, some white dwarfs that have not begun crystallising have clearly been observed to harbour a magnetic field.

Hence we conclude, similarly to \citet{Bagnulo2022,Manser2023,Elms2023}, that a crystallisation dynamo cannot be the unique mechanism to explain the magnetic nature of white dwarfs with masses $<$\,0.8\,\Msun. Recent theoretical studies have also raised doubts that a crystallisation dynamo is efficient enough to explain some or most of the magnetic white dwarfs, due to the small convective velocities, small kinetic energy flux reservoir and long rotation periods \citep{Ginzburg2022,Fuentes2023}. It could be coincidental that magnetic fields emerge in white dwarfs at a similar age range (1--3\,Gyr) to the onset of crystallisation, and that crystallisation does not explain the generation of magnetism in white dwarfs. These fields may have instead been present in their stars' red giant progenitors and emerged in the white dwarf.

The drop-off in magnetic frequency seen towards larger cooling ages in Fig.~\ref{fig:magneticfreq_age} is largely due to detection biases $-$ Zeeman splitting of spectral features is harder to detect at lower \Teff. The continuing effort to search for magnetism in DC white dwarfs using broadband filter polarimetry (e.g. \citealt{Berdyugin2022,Berdyugin2023}) provides vital information on whether the drop-off in magnetism as a function of cooling age at late ages is genuine. There may be an elusive population of very highly magnetic DA white dwarfs that have been misclassified due to their spectral features being so broadened and distorted that they resemble featureless DCs, which may further bias our results. There are also observational biases in the population of He-atmosphere white dwarfs, as those with metal lines enable sensitive detections of magnetic fields (\citealt{Hollands2017,Bagnulo2020}).

\subsection{Metal-Polluted White Dwarfs}
\label{sec:polluted}

\begin{figure}
    \centering
	\includegraphics[width=\columnwidth]{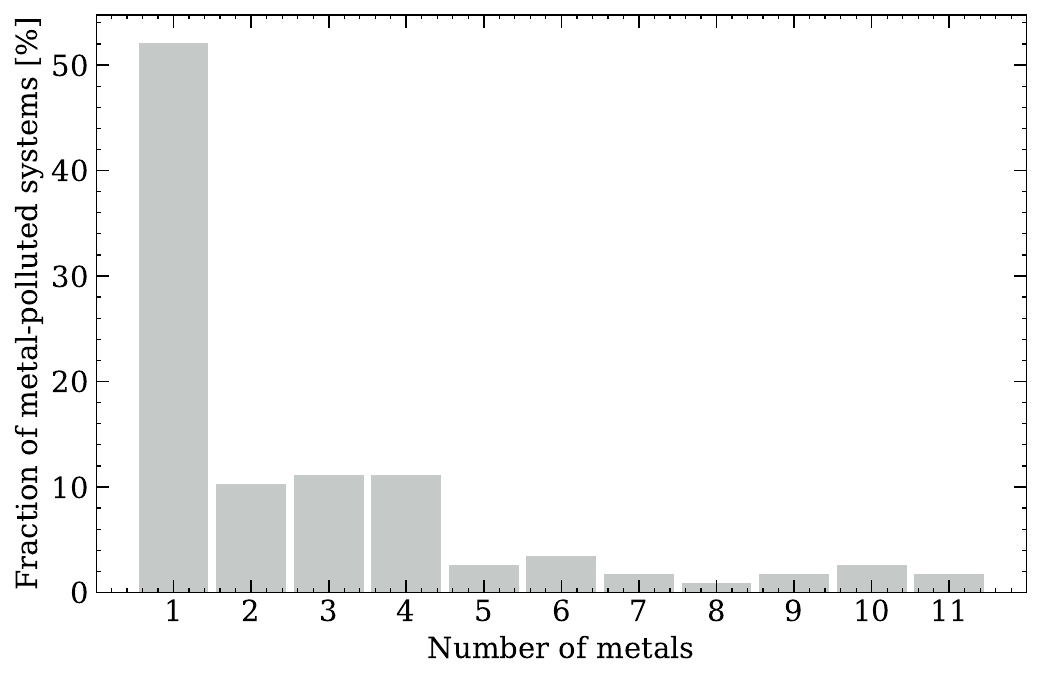}
	\caption{The fraction of metal-polluted white dwarfs in the 40 pc sample with different numbers of polluting metals. Data for this figure have been compiled from Williams et al. (in prep.).}
    \label{fig:num_metals}
\end{figure}

The pollution of white dwarf atmospheres by heavy elements is indicative of the accretion of planetary debris \citep{Zuckerman2007,Farihi2016,Veras2021}. Around 11\,per\,cent of white dwarfs in the 40\,pc sample are polluted. This fraction is dependent on spectroscopic resolution \citep{OBrien2023}, as well as both the \Teff\ range considered and the wavelength coverage (\citealt{Zuckerman2003,Koester2014}). Almost all metal-polluted white dwarfs in the 40\,pc sample only have optical spectra. In far-ultraviolet (FUV) spectroscopy, Si\,\textsc{ii} and C\,\textsc{ii} lines are prominent \citep{Koester2014}. In near-ultraviolet (NUV) spectroscopy, Mg\,\textsc{i} and Mg\,\textsc{ii} lines dominate \citep{Allard2018}. In cooler white dwarfs with \Teff\ $<$ 8000\,K, Ca generally has the greatest equivalent width for metals and in many cases is the only metal detected in optical spectra (see Fig.~\ref{fig:num_metals}). Therefore targeted high-resolution spectroscopic follow-up around the Ca\,\textsc{ii} H+K lines for 40\,pc white dwarfs should be carried out to improve completeness. We note that high-resolution observations do not always reveal more metals, many white dwarfs observed at very high resolution still only display Ca absorption features \citep{Zuckerman2003}.

\subsection{Gaia XP Spectra}

There are low-resolution \gaia\ BP/RP (commonly abbreviated to XP) spectra available for 99\,per\,cent of the 40\,pc white dwarf sample, which were released as part of \gaia\ DR3 \citep{Gaia2023}. By integrating under a \gaia\ XP spectrum convolved with the transmission of a desired photometric band, photometry can be generated in any arbitrary band that is within the wavelength coverage of \gaia\ \citep{Torres2023}. 

The $u - g$ colour from SDSS is well known to be sensitive to the Balmer jump for warm white dwarfs, which by its nature is only observed in white dwarfs with H-atmospheres. Therefore, an HR diagram using the $u - g$ colour allows for the separation of H- and He-atmosphere white dwarfs without the need for spectroscopy. 

We determine magnitudes from \gaia\ XP spectra by using the system defined in \citet{Holberg2006} for SDSS, and we differentiate our calculated magnitudes from catalogue SDSS magnitudes with the prime symbol ($'$). 

\begin{figure}
    \centering
	\includegraphics[width=\columnwidth]{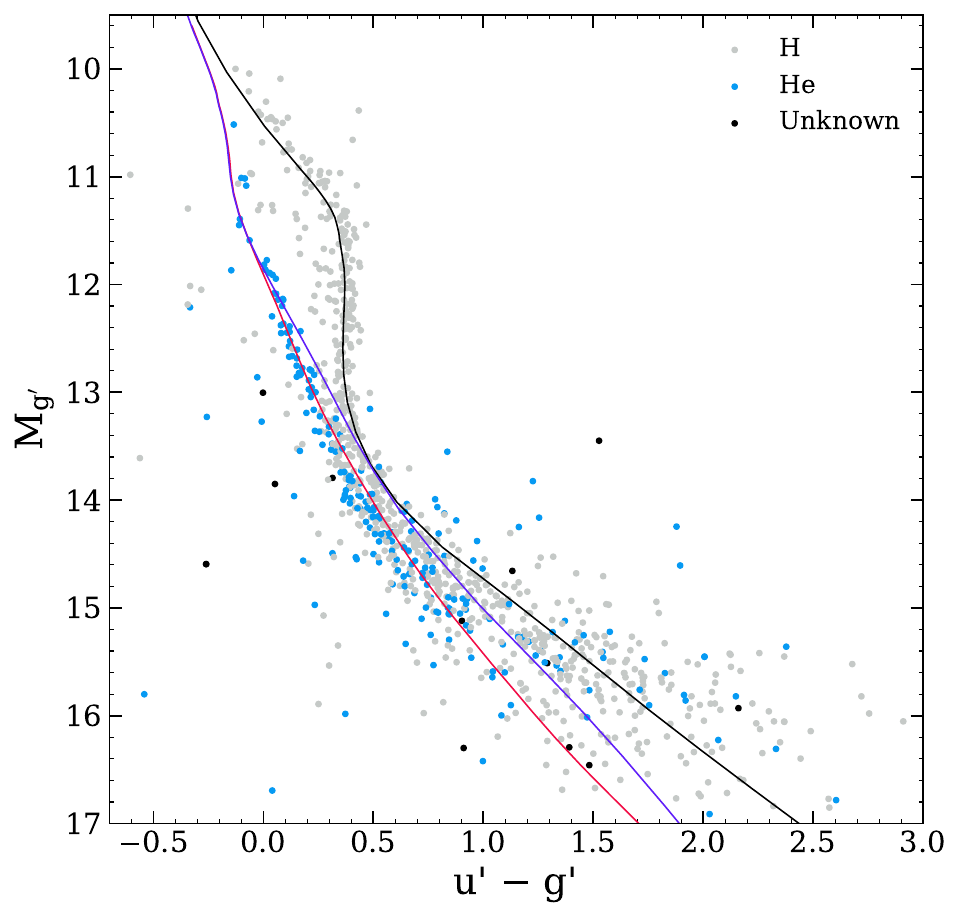}
	\caption{Photometry calculated using \gaia\ XP spectrophotometry in SDSS $u$ and $g$ bands, which we distinguish from catalogue SDSS photometry with a prime symbol. H-atmosphere white dwarfs are shown in grey, and He-atmosphere white dwarfs are in blue. Unconfirmed candidates are in black. The purple line indicates pure-He cooling tracks, the black line indicates pure-H cooling tracks and the red line indicates mixed H/He = $10^{-5}$ cooling tracks for a 0.6\,\Msun\ white dwarf.}
    \label{fig:SDSS_balmer}
\end{figure}

We demonstrate in Fig.~\ref{fig:SDSS_balmer} that integrating under \gaia\ XP spectra in $u'$ and $g'$ bands separates H- and He-atmospheres from the Balmer jump for bright white dwarfs ($M_{\rm g'} \lesssim 13$) within 40\,pc, as was previously found using catalogue SDSS photometry \citep[see, e.g.,][]{Caron2023}. However, Fig.~\ref{fig:SDSS_balmer} becomes noisy for fainter ($M_{\rm g'} \gtrsim 14$) white dwarfs, as the average S/N of the XP spectra becomes low, resulting in narrow-band $u'$ and $g'$ photometry becoming much less reliable than broadband $G_{\rm BP}$ and $G_{\rm RP}$ photometry. 

Most \gaia\ XP spectra are presented in the form of Gauss-Hermite polynomials. A truncation to the number of coefficients in these polynomials is often applied to fainter sources, as higher-order polynomials may attempt to fit noise \citep{Montegriffo2022}. There is a non-negligible effect in the resulting magnitudes when applying a truncation, as some signal is removed. This effect is particularly strong in the SDSS $u$-band, as cool white dwarfs are fainter in this wavelength region \citep[see also][]{Lopez2022b}. Therefore, we do not apply a truncation to any of the \gaia\ XP spectra when calculating photometry.

\begin{figure}
    \centering
	\includegraphics[width=\columnwidth]{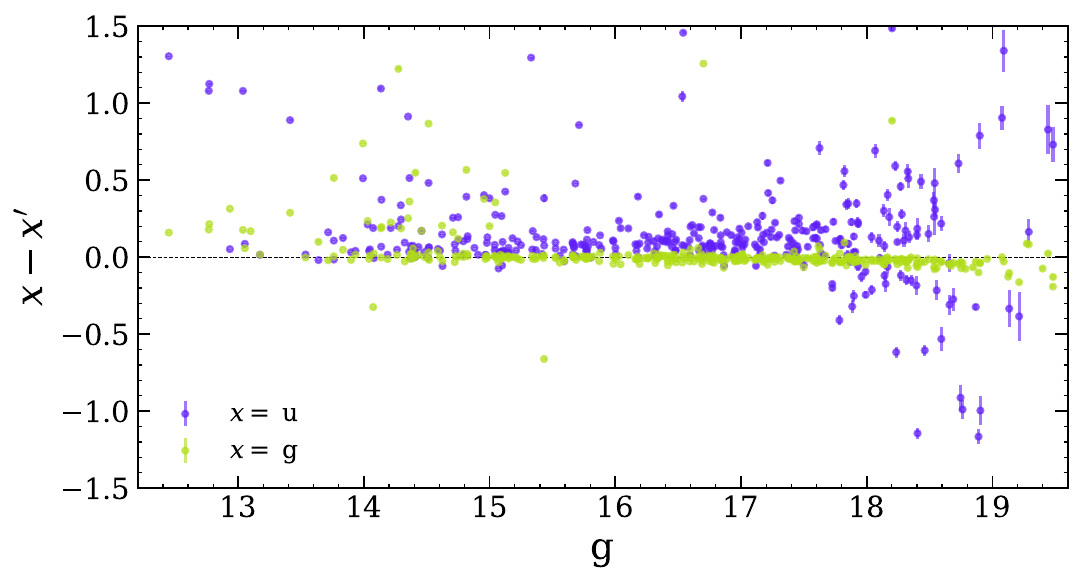}
	\caption{Comparison of catalogue SDSS photometry ($u$ and $g$) to photometry calculated using \gaia\ XP spectra in SDSS bands ($u'$ and $g'$). The difference is shown as a function of apparent SDSS $g$-magnitude.}
    \label{fig:SDSS_comparison}
\end{figure}

Fig.~\ref{fig:SDSS_comparison} demonstrates that \gaia\ XP spectra can accurately recreate SDSS $g$ colours down to faint absolute magnitudes, but the SDSS $u$ colour determination is problematic, and gets worse for fainter sources, which was also demonstrated by \citet{Vincent2023} for a much larger sample of white dwarfs. As a consequence, we make no attempt to test \gaia\ capabilities in identifying cool and faint DZ or DQ white dwarfs from $u'$, $g'$ or other ad-hoc narrow band filters. However, we note that \citet{Garcia-Zamora2023} and \citet{Vincent2023} have successfully used machine learning methods to classify white dwarfs into spectral types using \gaia\ XP spectra, which does not rely on the creation of narrow-band photometry.

\section{Conclusions}
\label{sec:conclusions}

We have presented the sample of white dwarfs within 40\,pc of the Sun, selected primarily from \gaia\ DR3, complete with spectroscopic follow-up. This is the largest volume-complete white dwarf sample to date, it is unaffected by reddening, and it will be the benchmark white dwarf sample for many years to come, until the era of multi-object spectroscopic surveys such as DESI \citep{DESI2022}, SDSS-V \citep{SDSS_dr16}, 4MOST \citep{4MOST}, and WEAVE \citep{WEAVE}. We find that white dwarfs make up $\approx$6\,per\,cent of stars in the local volume.

Our sample contains 1076 spectroscopically confirmed white dwarfs selected from the \gaia\ DR3 catalogue of \citet{Gentile2021}. Only 7 candidates from \citet{Gentile2021} within 40\,pc remain without spectroscopic follow-up. Spectroscopic observations have enabled us to determine the atmospheric composition of each white dwarf with \Teff\,$>$\,5000\,K (90\,per\,cent of the sample); therefore providing us with accurate \Teff\ and \logg\ from broadband \gaia\ photometric fitting. We present a spectroscopic analysis of two heavily polluted white dwarfs $-$ one well-known for decades and one newly discovered in the course of this work. Both stars are newly observed at high-resolution.

We perform a correction on the \gaia\ \Teff\ and masses of the sample to remove the effects of the low-mass issue in white dwarfs \citep{Hollands2018_Gaia,Mccleery2020,Bergeron2019,Cukanovaite2023}. The mass distribution of the sample reveals a main peak with a median mass of 0.61\,\Msun, with a shoulder (or secondary peak) of larger mass white dwarfs in the range 0.7--0.9\,\Msun. The spectroscopic heterogeneity of the sample prevents complete unbiased analyses based on spectral sub-types, such as metal pollution and magnetic white dwarfs. We observe no clear evidence of spectral evolution between H- and He-atmospheres at cool temperatures (\Teff\ $<$ 9000\,K).

We note that there are 28 suspected or confirmed white dwarfs within 40\,pc that did not make the cut of the \citet{Gentile2021} \gaia\ white dwarf catalogue. These are mostly in close binaries with main-sequence companions, such that their \gaia\ colours are blended. Constraining the numbers of these white dwarfs missing from the main selection will improve volume-completeness and inform future binary population models.

Comparison of the 40\,pc sample with the binary population models of \citet{Toonen2017} has demonstrated a distinct lack of wide double white dwarf binaries $-$ models predict almost a factor of ten more than we recover from our own search within 40\,pc. This deficit was observed in smaller volume-limited samples, but becomes more apparent within the 40\,pc volume. The numbers of other types of binary star systems agree well with models.

The space density of white dwarfs within 40\,pc of the Sun generally agrees with simulations from \citet{Cukanovaite2023}, but predictions from simulations depend on the vertical position of the Sun with respect to the Galactic plane and vertical scale height of the Galactic disk. 

We find that the mass distribution of magnetic white dwarfs has a primary peak centering around 0.7\,\Msun\ which is larger than the canonical non-magnetic white dwarf mass, and there is a hint of a secondary peak at very high masses ($M >$ 1.1\,\Msun). For white dwarfs with $M <$ 0.8\,\Msun, we find that the incidence of magnetism increases with cooling age, as was observed by \citet{Bagnulo2022}. Even the most relaxed constraints on core composition do not produce a good agreement between the increased incidence of magnetism and the onset of core crystallisation. Therefore, we argue that a dynamo generated by core crystallisation might not explain magnetic field generation in a significant fraction of magnetic white dwarfs.

\section*{Acknowledgements}
MOB, PET, and BTG received funding from the European Research Council under the European Union’s Horizon 2020 research and innovation programme numbers 101002408 (MOB and PET), and 101020057 (BTG). AB is a Postdoctoral Fellow of the Natural Sciences and Engineering Research Council of Canada. EC and PET were supported by grant ST/T000406/1 from the Science and Technology Facilities Council (STFC). TC was supported by a Leverhulme Trust Grant (ID RPG-2020-366). CM and BZ acknowledge support from US National Science Foundation grants SPG-1826583 and SPG-1826550. IP acknowledges a Warwick Astrophysics prize post-doctoral fellowship made possible thanks to a generous philanthropic donation. DK and MOB want to thank Simon Blouin for sharing his data on the high-density correction to the He-opacity, and carbon dredge-up sequences.

Some of the data presented herein were obtained at the W. M. Keck Observatory, which is operated as a scientific partnership among the California Institute of Technology, the University of California and the National Aeronautics and Space Administration. The Observatory was made possible by the generous financial support of the W. M. Keck Foundation. The authors wish to recognize and acknowledge the very significant cultural role and reverence that the summit of Maunakea has always had within the indigenous Hawaiian community. We are most fortunate to have the opportunity to conduct observations from this mountain. This paper includes data gathered with the 6.5-m Magellan Telescopes located at Las Campanas Observatory, Chile. Research at Lick Observatory is partially supported by a generous gift from Google. This work is based on observations collected at the European Southern Observatory under the ESO programme 111.24FZ.001.

This work has made use of data from the European Space Agency (ESA) mission \textit{Gaia} (\url{https://www.cosmos.esa.int/gaia}), processed by the \textit{Gaia} Data Processing and Analysis Consortium (DPAC, \url{https://www.cosmos.esa.int/web/gaia/dpac/consortium}). Funding for the DPAC has been provided by national institutions, in particular the institutions participating in the \textit{Gaia} Multilateral Agreement. This work has made use of the Python package GaiaXPy, developed and maintained by members of the Gaia DPAC, and in particular, Coordination Unit 5 (CU5), and the Data Processing Centre located at the Institute of Astronomy, Cambridge, UK (DPCI).

This research has made use of NASA’s Astrophysics Data System; the SIMBAD database, operated at CDS, Strasbourg, France; and the VizieR service. We use cooling models from \url{http://www.astro.umontreal.ca/~bergeron/CoolingModels}.

\section*{Data Availability}

Any reduced spectra from any spectrograph used in this article will be shared on reasonable request to the corresponding author.



\bibliographystyle{mnras}
\bibliography{mybib,pierbib}


\appendix

\section{Tables and Spectroscopy}

\begin{table*}
	\centering
        \caption{The catalogue of 1076 \gaia\ white dwarfs within 40\,pc, which is accessible online at this \href{https://cygnus.astro.warwick.ac.uk/phrtxn/}{link}. See Table~\ref{tab:online_table} for details.}
        \label{tab:all_online}
\end{table*} 

\begin{table*}
	\centering
        \caption{White dwarf candidates within 40\,pc in the \citet{Gentile2021} DR3 catalogue that do not have spectroscopic follow-up.}
        \label{tab:WDsnospectra}
        \begin{tabular}{llllll}
                \hline
                WD\,J Name & DR3 Source ID & Parallax & P$_{\rm WD}$ & \gaia\ \Teff & \gaia\ \logg \\
                \hline
                041359.12$-$212222.67 & 5090109228757394048 & 27.87 (0.08) & 0.66 & 4000 (50) & 7.11 (0.05) \\
                081219.58$-$261639.46 & 5694534861029977472 & 26.5 (0.8) & 0.30 & 5050 (20) & 6.88 (0.09) \\
                095953.92$-$502717.75 & 5405389966089801984 & 26.57 (0.04) & 1.00 & 6950 (140) & 8.12 (0.05) \\
                115007.08+240403.54 & 4004185576130620288 & 33.2 (0.3) & 1.00 & 3600 (200) & 8.5 (0.1) \\
                $*$224600.88$-$060947.02 & 2611515835965491968 & 27 (1) & 0.79 & -- & -- \\
                \hline
        \end{tabular}\\
        $*$: This system is likely to be a brown dwarf (see Section~\ref{sec:missing} for details). \\
\end{table*} 

\begin{table*}
	\centering
        \caption{Confirmed white dwarfs and candidates in the \citet{Gentile2021} DR3 catalogue that are within 1$\sigma_\varpi$ of 40\,pc.}
        \label{tab:onesigma}
        \begin{tabular}{llllll}
                \hline
                WD\,J Name & DR3 Source ID & Parallax & P$_{\rm WD}$ & SpT & Reference \\
                \hline
                014240.09$-$171410.85 & 5142336825646176256 & 24.97 (0.09) & 1.00 & DAH & \citet{OBrien2023} \\
                054858.25$-$750745.20 & 4648527839871194880 & 24.97 (0.09) & -- & DZH & \citet{OBrien2023} \\
                055231.03+164250.27 & 3349849778193723008 & 24.97 (0.04) & 1.00 & DBA & \citet{Tremblay2020} \\
                080247.02+564640.62 & 1081514379072280320 & 24.9 (0.2) & 0.99 & DC & \citet{Tremblay2020} \\
                100819.19+121813.94 & 3881550619014086912 & 23 (3) & 1.00 & -- & -- \\
                $*$102834.88$-$000029.39 & 3831059120921201280 & 24.997 (0.026) & 0.03 & DA+M & \citet{Gianninas2011} \\
                122257.77$-$742707.70 & 5838312052354944256 & 24.97 (0.07) & 0.99 & DA & \citet{OBrien2023} \\
                133340.50$-$370550.65 & 6162813873991704960	& 24.99 (0.07) & 0.99 & -- & -- \\
                134118.69+022737.01 & 3713218786120541824 & 24.96 (0.09) & 0.99 & DQ & \citet{Kilic2010} \\
                180218.60+135405.46 & 4496751667093478016 & 25.00 (0.04) & 0.99 & DAZ & \citet{Tremblay2020} \\
                183010.48$-$244209.53 & 4077104740685645056 & 24 (1) & 1.00 & -- & -- \\
                193500.68$-$172443.11 & 4180014832789446400 & 25.0 (0.2) & 1.00 & DC & \citet{Tremblay2020} \\ 
                193501.33$-$072527.42 & 4207055367062840320 & 24.9 (0.2) & 0.96 & DC & This work \\
                214810.74$-$562613.14 & 6460523071166427392 & 24.98 (0.08) & 1.00 & DAH & \citet{OBrien2023} \\
                222919.46$-$444138.86 & 6520516480027596288 & 24.97 (0.03) & 1.00 & DA & \citet{Beers1992} \\
                \hline
        \end{tabular}\\
        $*$: This system has a low P$_{\rm WD}$ due to the M-dwarf companion. \\
\end{table*} 

\begin{table*}
	\centering
        \caption{White dwarfs within 40\,pc that are not in the \citet{Gentile2021} DR3 catalogue.}
        \label{tab:missingWDs}
        \begin{tabular}{llllll}
                \hline
                DR3 Source ID & WD Name & Parallax & SpT & SpT Reference & Note \\
                \hline
                -- & Procyon B & 285 (1) & DQZ+F & \citet{Limoges2015} & (1) \\
                1355264565043431040	& WD\,1708+437 & 131.6 (0.4) & WD+M & \citet{Delfosse1999} & (2) \\
                4937000898856154624 & WD\,0210$-$510 & 92.6 (0.1) & DQ & \citet{Farihi2013} & (3) \\
                975968340910692608 & WD\,0727+482B & 88.72 (0.03) & DA & \citet{Limoges2015} & (4) \\
                975968340912517248 & WD\,0727+482A & 88.72 (0.03) & DA & \citet{Limoges2015} & (4) \\
                3223516063958808064	& GJ\,207.1 & 63.36 (0.05) & WD+M & \citet{Baroch2021} & (2) \\
                1005873614080407296 & LHS\,1817 & 61.43 (0.05) & WD+M & \citet{Winters2020} & (2) \\
                2185710338703934976 & WD\,2003+542 & 60.30 (0.03) & WD+M & \citet{Gizis1998} & (2) \\
                4788741548375134336 & WD\,0419$-$487 & 47.2 (0.02) & DA+M & \citet{Gianninas2011} & (2) \\
                1362295082910131200 & HD\,159062B & 46.19 (0.01) & WD+G & \citet{Hirsch2019} & (4) \\
                2274076301516712704 & WD\,2126+734B & 44.91 (0.07) & DC & \citet{Zuckerman1997} & (3) \\
                6431977687725247104	& SCR\,J1848$-$6855 & 43.9 (0.1) & WD+M & \citet{Jao2014} & (2) \\
                3701290326205270528 & WD\,1214+032 & 42.77 (0.04) & DA & \citet{Limoges2015} & (3) \\
                2983256662868370048	& GJ\,3346 B & 42.24	(0.04) & WD+K & \citet{Bonavita2020} & (3) \\
                -- & Regulus B & 41.1 (0.4) & WD+B & \citet{Gies2020} & (1) \\
                3729017810434416128	& HD\,114174 B & 37.87 (0.02) & WD+G & \citet{Crepp2013} & (2) \\
                1548104507825815296 & WD\,1213+528 & 34.95 (0.02) & DA+M & \citet{Limoges2015} &(2) \\
                1550299304833675392 & WD\,1324+458 & 32.77 (0.02) & DA+M & \citet{vandenBesselaar2007} & (2) \\
                5389590533737966208 & WD\,1108$-$408 & 32.5 (0.2) & DC & \citet{Monteiro2006} & (3) \\
                6665685378201412992 & CD$-$53\,8345B & 31.3 (0.08) & DA & \citet{OBrien2023} & (3) \\
                4478524169500496000 & HD\,169889 & 28.27 (0.03) & WD+G & \citet{Crepp2018} & (2)\\
                -- & WD\,1634$-$573 & 27.9 (0.2) & DOA+K & \citet{Dreizler1996} & (1) \\
                1962707287281651712 & PM\,J22105+4532 & 27.76 (0.09) & DC & \citet{Limoges2013} & (3) \\
                2643862402903084544	& 12 Psc B & 27.53 (0.02) & WD+G & \citet{Bowler2021} & (2) \\
                3845263368043086080 & WD\,0911+023 & 27.1 (0.6) & WD+B & \citet{Holberg2013} & (2) \\
                3817534337626005632 & WD\,1120+073 & 27.0 (0.5) & DC & \citet{Limoges2015} & (4) \\
                759601941671398272 & WD\,1133+358 & 25.91 (0.06) & DC+M & \citet{Putney1997} & (2) \\
                3000597125173673088 & PM\,J06157$-$1247 & 25.8 (0.1) & WD+M & \citet{FajardoAcosta2016} & (2) \\
                \hline
        \end{tabular}\\
        Notes: (1) Missing entirely from \gaia\ DR3, (2) Unresolved white dwarf plus main sequence binary, (3) Missing or incorrect colours, (4) Missing five-parameter astrometry. For many binaries, parallaxes are of the companion. Where no \gaia\ ID is available, parallaxes are from \textit{Hipparcos} \citep{vanLeeuwen2007}. A white dwarf that has not been spectroscopically confirmed is denoted as WD.\\  
\end{table*}   

\begin{table*}
	\centering
        \caption{White dwarfs within 40\,pc with unreliable \gaia\ parameters. Best-fit parameters are instead taken from literature.}
        \label{tab:bad_params}
        \begin{tabular}{lllllll}
                \hline
                WD\,J Name & Parallax & SpT & \Teff & Mass & Description & Reference \\
                 & [mas] &  & [K] & [\Msun] & & \\
                \hline 
                021348.83$-$334530.03 & 53.33 (0.06) & DAZ & 5150 (150) & 0.4 (0.1) & (1) & This work \\
                023538.55$-$303225.52 & 30.6 (0.2) & DC & -- & -- & (2) & This work \\
                022432.27$-$285459.46 & 34.5 (0.1) & DC & 4880 (160) & 1.071 (0.003) & (2) & \citet{Bergeron2022} \\
                034646.52+245602.67 & 25.3 (0.2) & DC & 3640 (60) & 0.423 (0.007) & (2) & \citet{Bergeron2022} \\
                050600.41+590326.89 & 27.7 (0.3) & DC & -- & -- & (2) & This work \\
                063038.60$-$020550.49 & 46.72 (0.03) & DA+M & 6910 (140) & 0.53 (0.1) & (1) & \citet{Gianninas2011} \\
                064509.30$-$164300.72 & 374.5 (0.2) & DA & 25\,970 (380) & 0.98 (0.03)  & (1) & \citet{Giammichele2012} \\
                075508.95$-$144550.95 & 25.55 (0.02) & DA+M & 19\,440 (290) & 0.58 (0.02) & (1) & \citet{Gianninas2011} \\
                085357.69$-$244656.23 & 38.8 (0.1) & DC & 3740 (40) & 0.672 (0.003) & (2) & \citet{Bergeron2022} \\
                090208.37+201051.57 & 26.1 (0.1) & DQ & 5500 (110) & 0.71 (0.01) & (3) & \citet{Blouin2019c} \\
                101141.58+284559.07 & 67.88 (0.06) & DQpecH & 4340 (170) & 0.70 (0.06) & (3) & \citet{Blouin2019c} \\
                104410.24$-$691818.08 & 34.18 (0.02) & DA+M & 22\,570 (330) &  0.54 (0.02) & (1) & \citet{Gianninas2011} \\
                110217.52+411321.18 & 28.7 (0.3) & DC & 3790 (20) & 0.56 (0.01) & (2) & \citet{Caron2023} \\
                122048.70+091413.08 & 26.7 (0.3) & DC & 3890 (60) & 1.081 (0.008) & (2) & \citet{Bergeron2022} \\
                130503.44+702243.05 & 28.9 (0.2) & DC & -- & -- & (2) & \citet{Tremblay2020} \\
                140324.75+453333.02 & 29.1 (0.2) & DC & 4820 (20) & 1.184 (0.003) & (2) & \citet{Bergeron2022} \\
                155647.51$-$080601.24 & 30.6 (0.2) & DC & 4880 (110) & 1.054 (0.004) & (2) & \citet{Bergeron2022} \\
                165401.26+625354.91 & 32.5 (0.1) & DC & 4990 (30) & 1.049 (0.002) & (2) & \citet{Bergeron2022} \\
                192206.20+023313.29 & 25.4 (0.3) & DZ & 3340 (50) & 0.57 (0.03) & (2) & \citet{Elms2022} \\
                195211.78$-$732235.48 & 31.2 (0.3) & DC & -- & -- & (2) & \citet{OBrien2023} \\
                201231.78$-$595651.67 & 60.80 (0.03) & DC & 4910 (210) & 0.44 (0.01) & (2) & \citet{Giammichele2012} \\
                214756.59$-$403527.79 & 35.8 (0.5) & DZQH & 3050 (40) & 0.69 (0.02) & (4) & \citet{Elms2022} \\
                215406.45$-$011709.55 & 39.2 (0.1) & DA+M & 9190 (130) & 0.58 (0.03) & (1) & \citet{Giammichele2012} \\
                230550.09+392232.88 & 27.9 (0.1) & DC & 4550 (30) & 0.698 (0.004) & (2) & \citet{Bergeron2022} \\
                231732.63$-$460816.77 & 26.0 (0.3) & DQpec & 4080 (100) & 0.70 (0.01) & (1) & This work \\
                \hline 
        \end{tabular}\\
        Notes: (1) \gaia\ photometry contaminated by nearby main-sequence star or unresolved main-sequence companion, (2) IR-faint white dwarf with no unique \gaia\ fit solution, (3) Strong C$_2$ molecular features affect \gaia\ photometry, (4) Ultra-cool white dwarf. \\ 
\end{table*}

\begin{table*}
	\centering
        \caption{List of wide binaries and higher-order systems, where at least one companion is a white dwarf, within 40\,pc. The table is accessible online at this \href{https://cygnus.astro.warwick.ac.uk/phrtxn/}{link}. The projected separation between sources is given in pc and au, and the tangential velocity difference $delta\_vtan$ is in km/s.}
        \label{tab:widebinaries_online}
\end{table*} 

\begin{table*}
	\centering
        \caption{Abundances and upper limits of elements determined from combined fitting of spectra and photometry with \citet{Koester2010} models.}
        \label{tab:metal_abundances}
        \begin{tabular}{llllll}
                \hline
                Parameter & WD\,J0213$-$3345 & WD\,J1154$-$6239 & WD\,J1927$-$0355 & WD\,J2141$-$3300 & WD\,J2317$-$4608 \\
                \hline
                \Teff\ [K] & 5150 (150) & 5100 (100) & 6540 (150) & 6870 (150) & 4075 (100) \\
                \logg\ [cm s$^{-2}$] & 7.6 (0.2) & 7.97 (0.05) & 7.99 (0.04) & 7.96 (0.04) & 8.2 (0.2) \\
                Composition, X & H & H & He & He & He \\
                log(H/X) & -- & -- & $-$3.5 & $-$3.5 & -- \\
                log(C/X) & -- & -- & -- & -- & $-$8.3 (1.0) \\
                log(Na/X) & -- & -- & $<-$9.4& $-$9.2 (0.2) & -- \\
                log(Mg/X) & -- & -- & $-$7.00 (0.15) & $-$7.50 (0.15) & -- \\
                log(Al/X) & -- & -- & $<-$8.6 & $-$8.5 (0.3) & -- \\
                log(Si/X) & -- & -- & $<-$7.6 & $-$7.2 (0.2) & -- \\
                log(Ca/X) & $-$10.9 (0.1) & $-$11.1 (0.2) & $-$9.1 (0.11) & $-$8.9 (0.1) & -- \\
                log(Sc/X) & -- & -- & $<-$11.7 & $<-$11.7 & -- \\
                log(Ti/X) & -- & -- & $-$10.7 (0.1) & $-$10.0  (0.1) & -- \\
                log(V/X) & -- & -- & $<-$10.5 & $<-$10.4 & -- \\
                log(Cr/X) & -- & -- & $-$10.2 (0.2) & $-$10.0 (0.2) & -- \\
                log(Mn/X) & -- & -- & $<-$9.2 & $<-$9.3 & -- \\
                log(Fe/X) & -- & -- & $-$8.0 (0.1) & $-$8.2 (0.1) & -- \\
                log(Co/X) & -- & -- & $<-$11.0 & $<-$10.7 & -- \\
                log(Ni/X) & -- & -- & $-$9.3 (0.2) & $-$9.2 (0.2) & -- \\
                log(Cu/X) & -- & -- & $<-$11.7 & $<-$11.7 & -- \\
                log(Sr/X) & -- & -- & $<-$12.3 & $-$12.1 (0.3) & -- \\
                log(Ba/X) & -- & -- & $<-$12.2 & $<-$12.2 & -- \\
                \hline
        \end{tabular}\\
\end{table*} 

\begin{figure*}
    \centering
	\includegraphics[width=\textwidth]{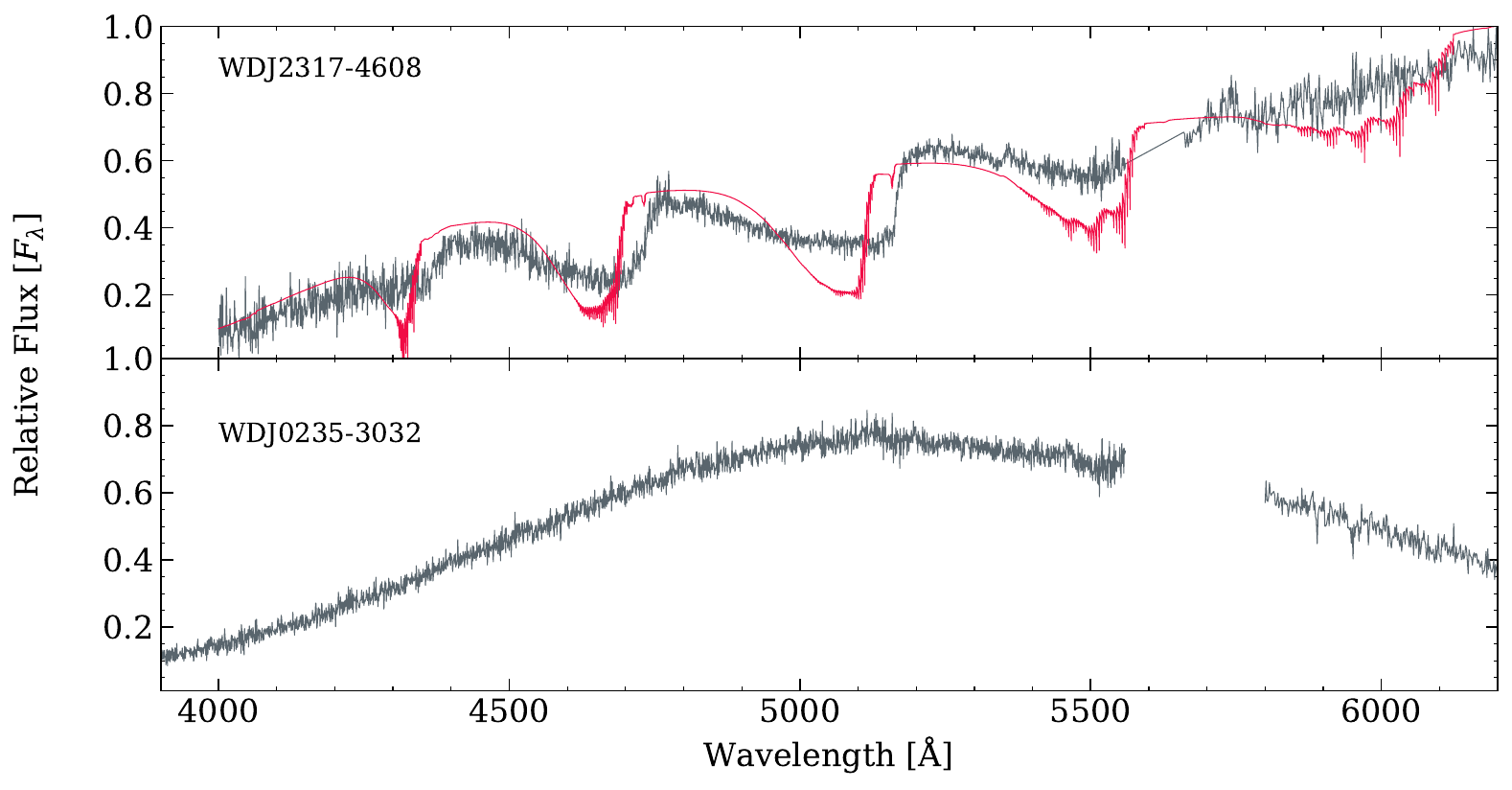}
	\caption{Spectra of two white dwarfs observed with X-Shooter. Top: the DQpec white dwarf \textbf{WD\,J2317$-$4608}, with our best-fit model spectrum over-plotted in red, based on DQ models \citep{Koester2010}. Bottom: the IR-faint DC white dwarf \textbf{WD\,J02353$-$30322}.}
    \label{fig:dq_XShooter}
\end{figure*}

\begin{figure*}
    \centering
	\includegraphics[width=\textwidth]{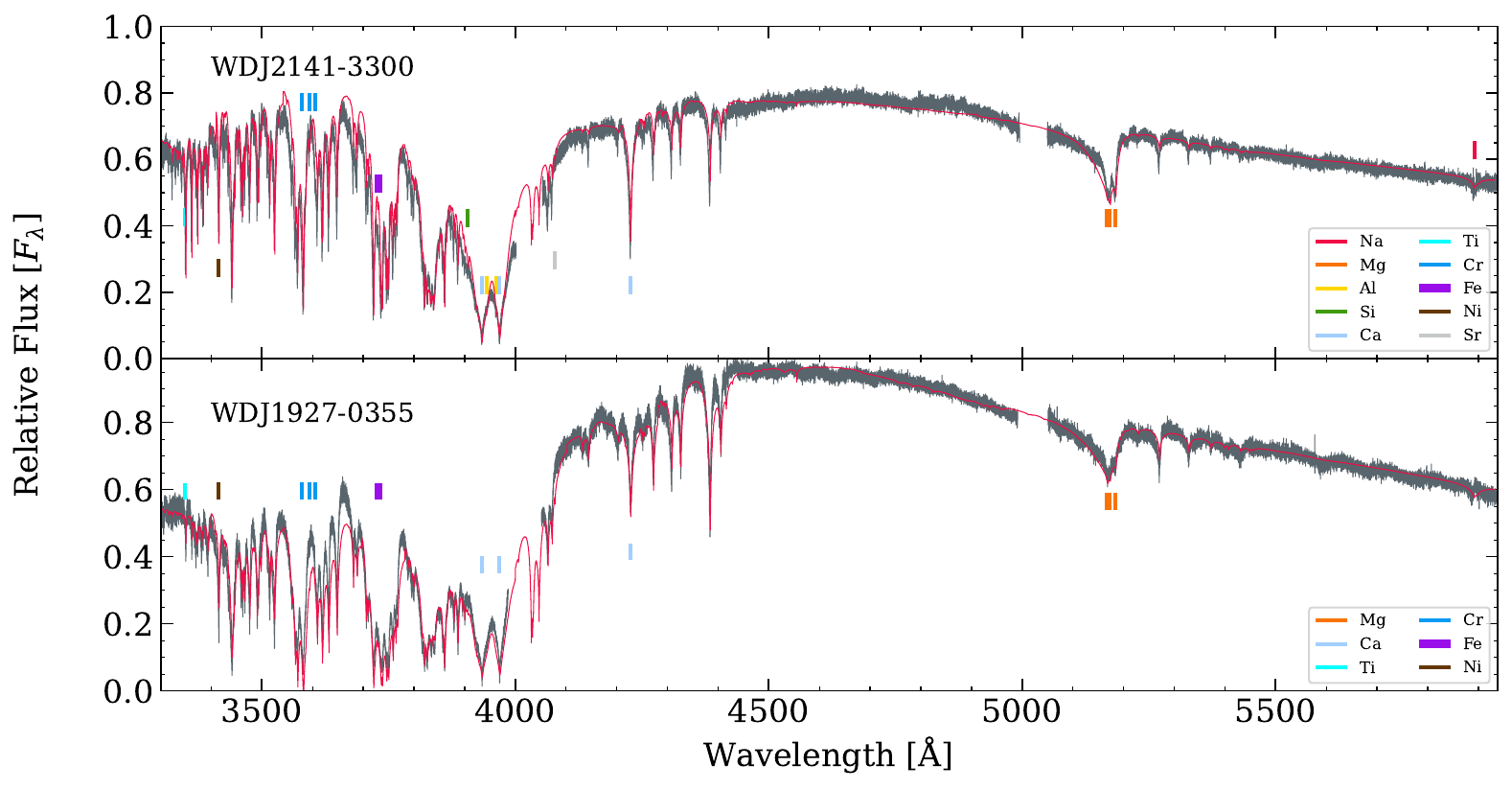}
	\caption{Spectra of two metal-polluted white dwarfs, \textbf{WD\,J2141$-$3300} and \textbf{WD\,J1927$-$0355}, observed with HIRES. A few prominent metal features are highlighted with ticks. Our best-fit model spectra are over-plotted in red \citep{Koester2010}.}
    \label{fig:dz_HIRES_blue}
\end{figure*}

\begin{figure}
    \centering
	\includegraphics[width=\columnwidth]{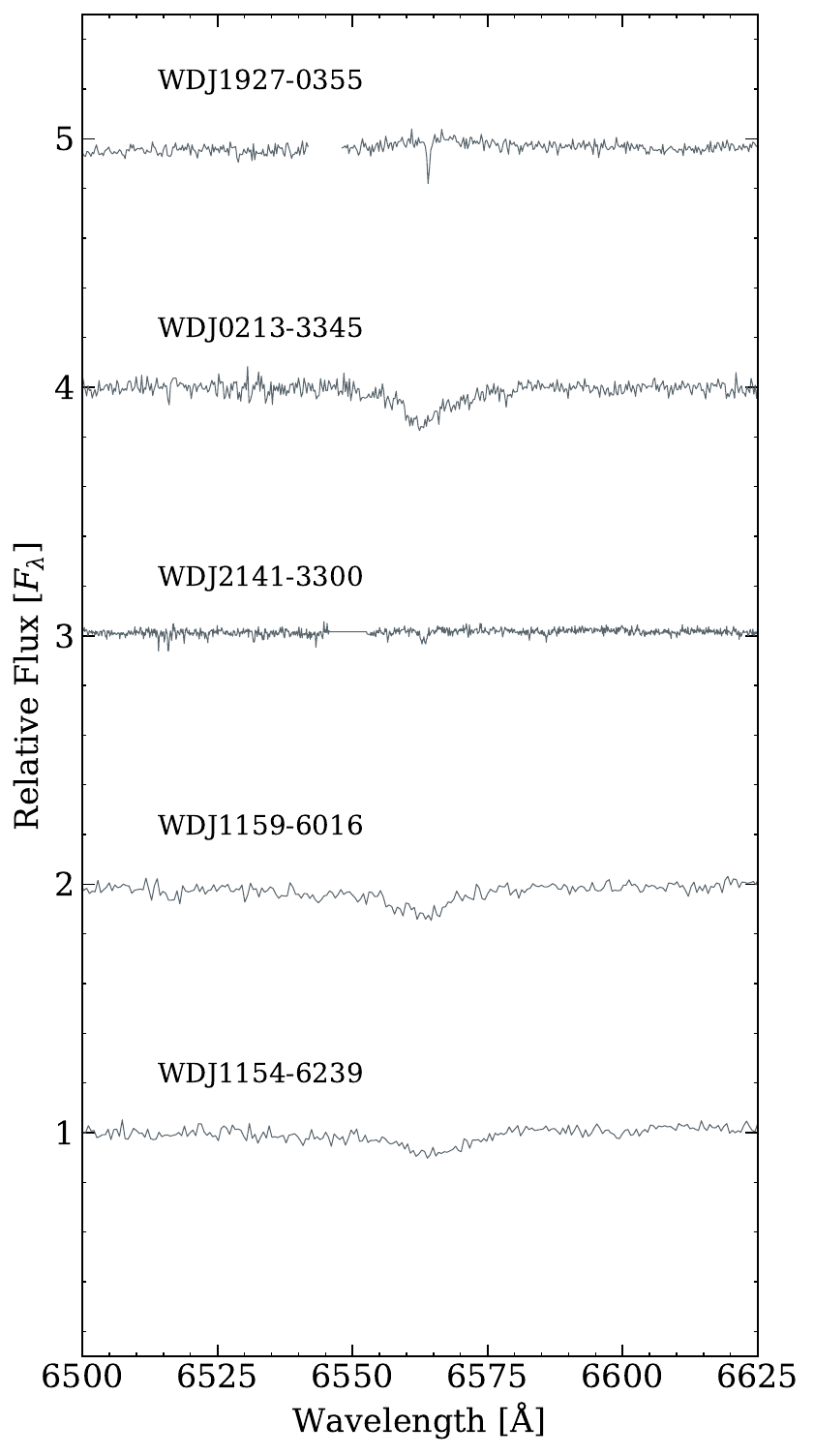}
	\caption{H$\alpha$-line spectra from HIRES (first and second), MIKE (third) and MagE (fourth and fifth) observations of \textbf{WD\,J1927$-$0355}, \textbf{WD\,J2141$-$3300}, \textbf{WD\,J0213$-$3345}, \textbf{WD\,J1159$-$6016} and \textbf{WD\,J1154$-$6239}. Spectra are normalised to unity and then offset by 1.}
    \label{fig:da_MagE_MIKE_HIRES}
\end{figure}

\begin{figure}
    \centering
	\includegraphics[width=\columnwidth]{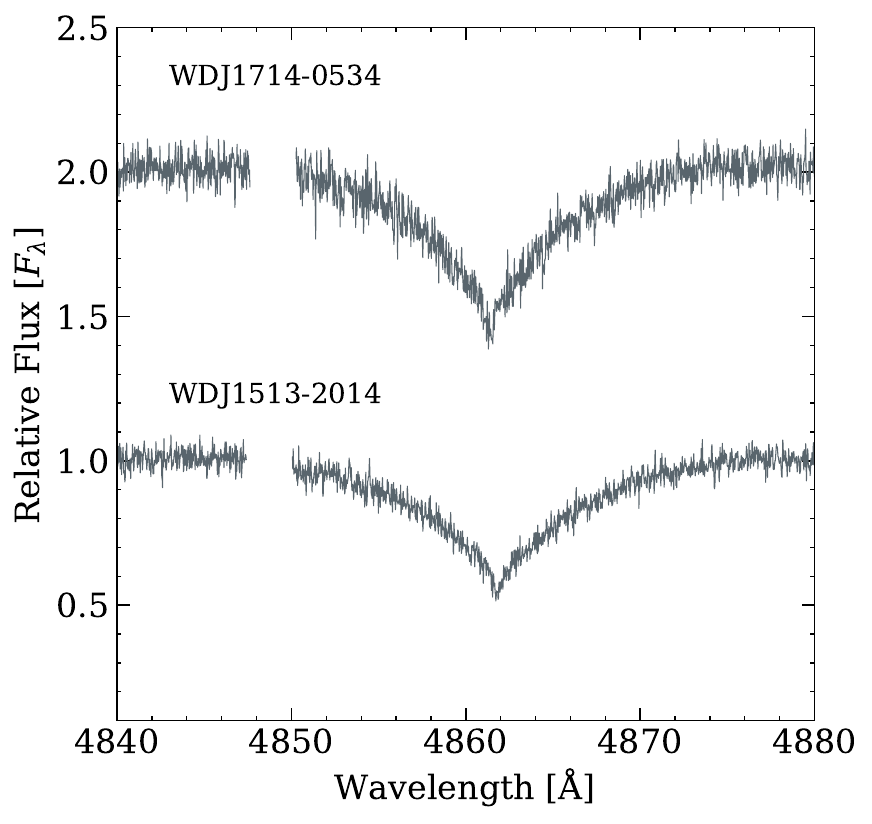}
	\caption{H$\beta$-line spectra from HIRES observations of \textbf{WD\,J1513$-$2014} and \textbf{WD\,1714$-$0534}.}
    \label{fig:da_HIRES}
\end{figure}

\begin{figure}
    \centering
	\includegraphics[width=\columnwidth]{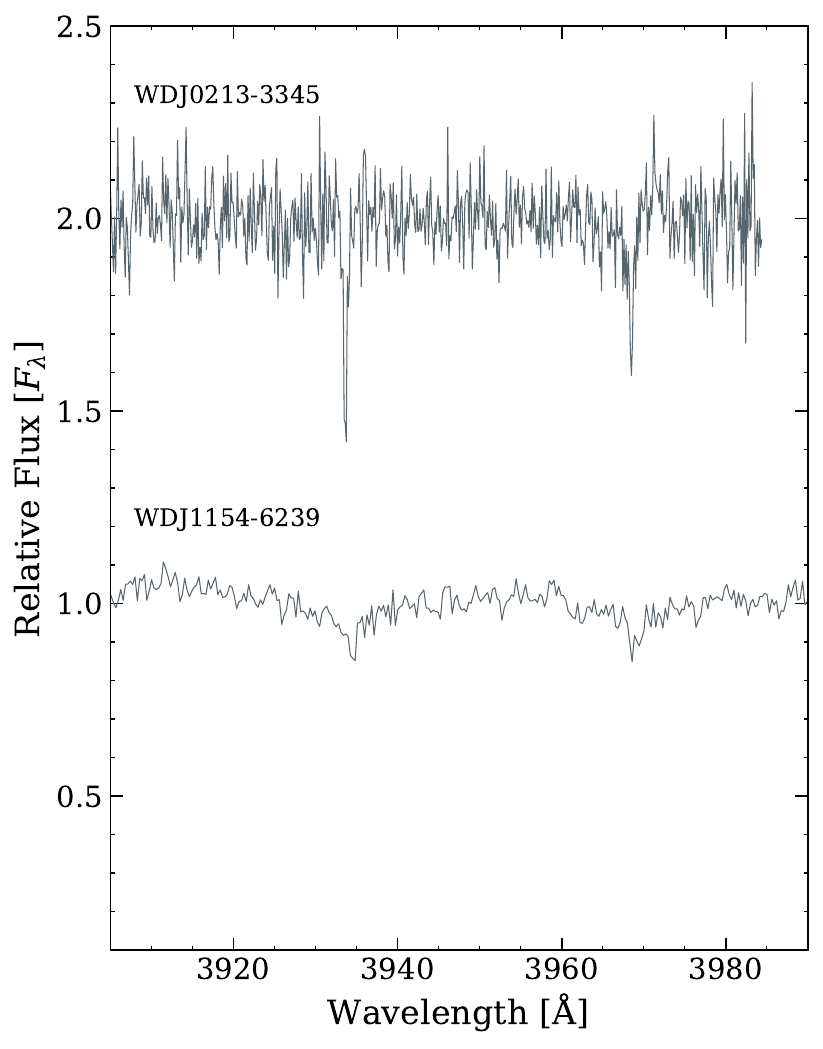}
	\caption{Ca\,\textsc{ii} H+K-line spectra from MIKE (top) and MagE (bottom) observations of \textbf{WD\,J0213$-$3345} and \textbf{WD\,J1154$-$6239}. Both spectra are of DAZ white dwarfs with intrinsic calcium absorption.}
    \label{fig:daz_calcium}
\end{figure}

\begin{figure}
    \centering
	\includegraphics[width=\columnwidth]{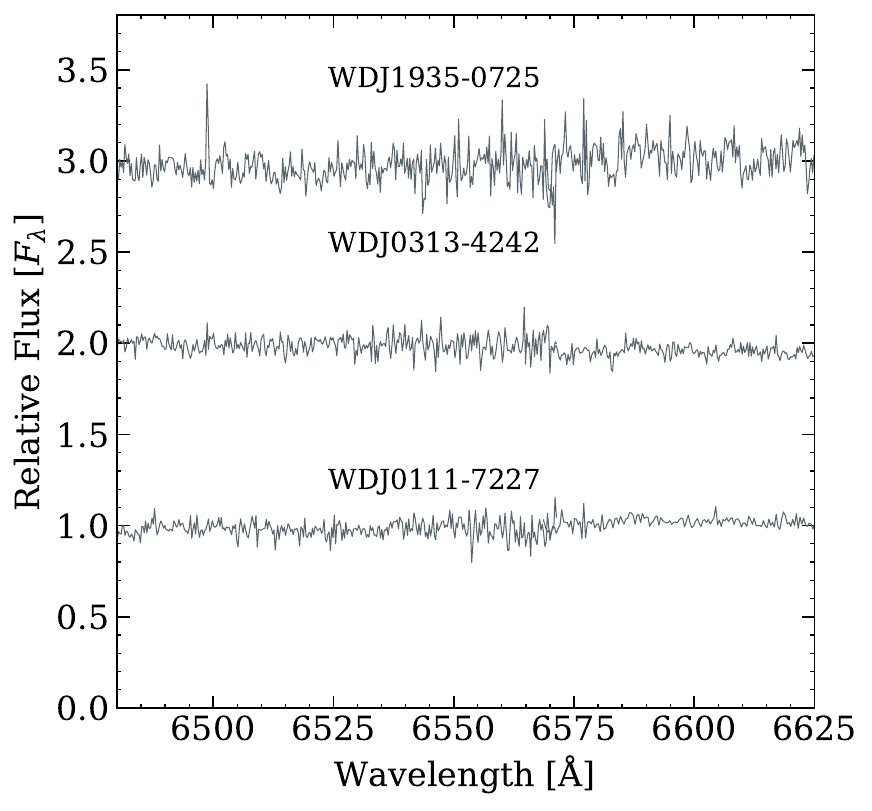}
	\caption{Spectra from MIKE of \textbf{WD\,J1935$-$0725}, \textbf{WD\,0313$-$4242}, and \textbf{WD\,0111$-$7227}, which are all DC white dwarfs showing no hydrogen features around the H$\alpha$ region.}
    \label{fig:dc_mike}
\end{figure}

\begin{figure*}
    \includegraphics[viewport= 1 20 520 720, scale=0.85]{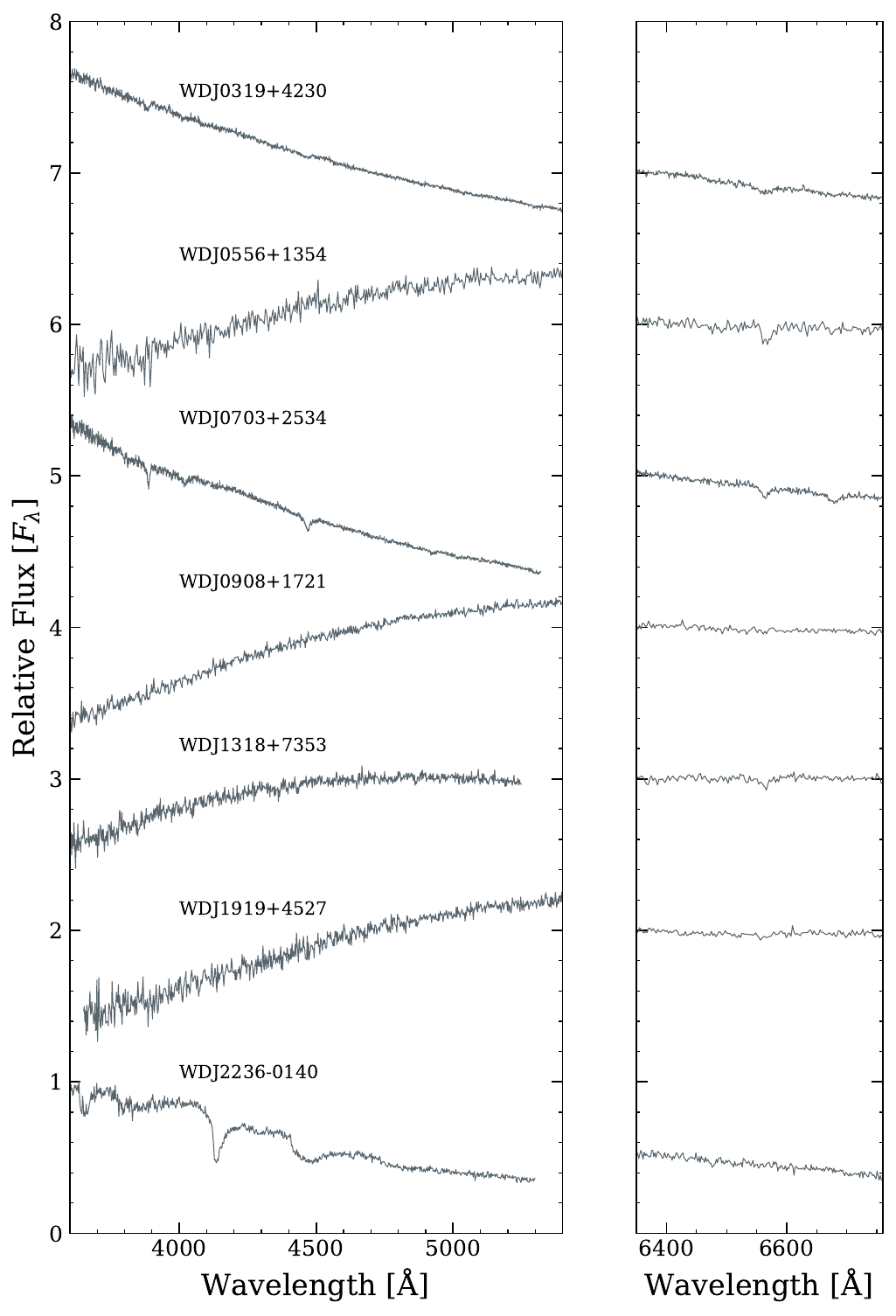}
    \vspace*{10mm}
    \caption{Spectra of white dwarfs observed with Kast.}
    \label{fig:kast}
\end{figure*}

\begin{figure*}
    \includegraphics[viewport= 1 20 520 720, scale=0.85]{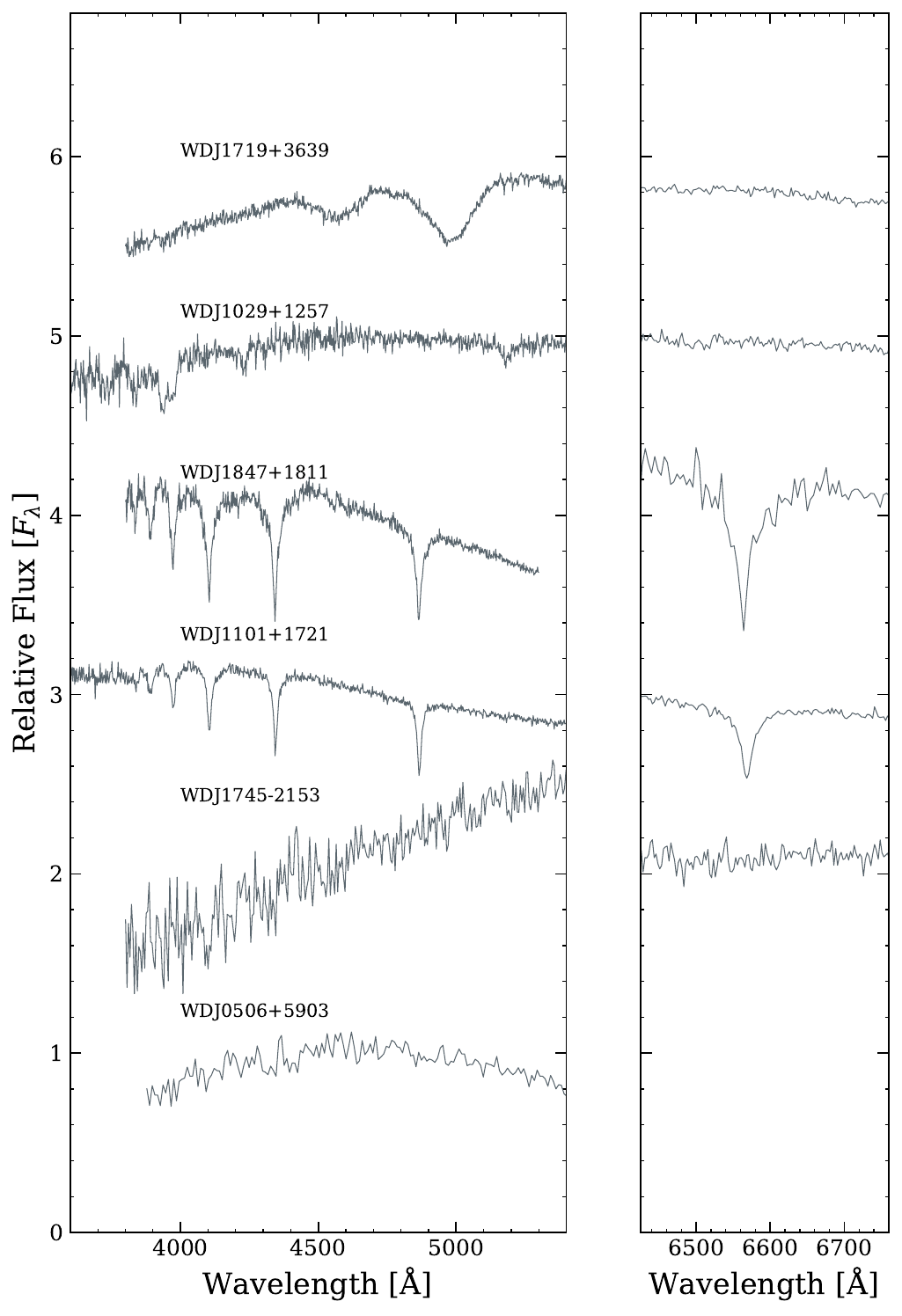}
    \vspace*{10mm}
    \caption{Spectra of white dwarfs observed with Kast.}
    \label{fig:kast2}
\end{figure*}

\bsp	
\label{lastpage}
\end{document}